%% file: main.tex
\documentclass[conference]{IEEEtran}
\IEEEoverridecommandlockouts
\usepackage{amsmath,amssymb,amsfonts}
\usepackage{graphicx}
\usepackage{pgfplots}
\usepackage{pgfplotstable}
\usepackage{tikz}
\usepackage{textcomp}
\usepackage{xcolor}
\usepackage{fancyhdr}
\usepackage[hyphens]{url}
\usepackage{color}
\usepackage{listings}
\lstdefinestyle{codeside}{%
  basicstyle=\ttfamily\fontsize{5.25}{6.1}\selectfont,
  tabsize=2,
  frame=single,
  framesep=2pt,
  framerule=0.3pt,
  aboveskip=0pt,
  belowskip=0pt,
  columns=fullflexible,
  keepspaces=true,
  breaklines=true,
  xleftmargin=0.55em,
  framexleftmargin=0.55em,
}
\newcommand{\codepairstart}{%
  \codepairstartasym{0.485}{0.485}%
}
\newcommand{\codepairstartasym}[2]{%
  \noindent\begin{tabular}[t]{@{}p{#1\columnwidth}@{\hspace{0.015\columnwidth}}p{#2\columnwidth}@{}}%
  \begin{minipage}[t]{\linewidth}%
}
\newcommand{\codepairmid}{%
  \end{minipage} & \begin{minipage}[t]{\linewidth}%
}
\newcommand{\codepairend}{%
  \end{minipage} \\ \end{tabular}%
}
\usepackage{float}
\usepackage[noabbrev]{cleveref}
\usepackage{dblfloatfix}
\usepackage{xr}
\usepackage{scalefnt}
\usepackage{framed}
\usetikzlibrary{matrix}
\usetikzlibrary{patterns}
\usetikzlibrary{arrows.meta,calc}
\usepackage{array}
\usepackage{tabularray}
\usepackage{booktabs}
\usepackage{titlesec}
\usepackage{flexisym}
\usepackage[font=small,skip=0pt]{caption}
\usepackage{enumitem}
\usepackage{algorithm} 
\usepackage{algpseudocode} 
\usepackage{subcaption}
\usepackage[citestyle=numeric-comp,bibstyle=ieee,sorting=none,backend=biber]{biblatex}

\titlespacing*{\subsubsection}{0pt}{0.35\baselineskip plus 0.1\baselineskip minus 0.1\baselineskip}{0.15\baselineskip}

\def\BibTeX{{\rm B\kern-.05em{\sc i\kern-.025em b}\kern-.08em
    T\kern-.1667em\lower.7ex\hbox{E}\kern-.125emX}}

\begin{document}

\title{The Fallacy of Independent Ceilings:\\
Characterizing Coupled Load-Branch Stall Interaction\\
}

\author{\IEEEauthorblockN{Matthew Constant\textsuperscript{1}\thanks{This work has been completed when Matthew was at University of Rhode Island.} and Resit Sendag\textsuperscript{2}}
\IEEEauthorblockA{\textsuperscript{1}AMD\quad \textsuperscript{2}Department of Electrical and Computer Engineering, University of Rhode Island \\
matthew.constant@amd.com,\quad sendag@uri.edu}
}

\maketitle

\begin{abstract}
\input{abstract}
\end{abstract}


\section{Introduction}
\input{introduction}

\section{Background and Related Work}
\label{sec:related}
\input{related_work}

\section{Motivation}
\label{sec:motivation}
\input{motivation}

\section{Methodology}
\label{sec:methodology}
\input{methodology}

\section{Results}
\label{sec:results}
\input{analysis}

\section{Discussion}
\label{sec:discussion}
\input{discussion}

\section{Conclusion}
\label{sec:conclusion}
\input{Conclusion}

\clearpage
\printbibliography


\end{document}

%% file: abstract.tex
Branch mispredictions and data-cache misses are usually evaluated as separate bottlenecks: studies report perfect-branch or perfect-cache speedups as isolated upper bounds and often treat their product as the joint ceiling. In irregular workloads, however, hard-to-predict branches and cache-missing loads often appear in the same hot loops. Removing only one penalty can expose the other: faster memory lets the core reach mispredicted branches sooner, while better branch prediction leaves more long-latency loads occupying the out-of-order window. We call this interaction \emph{symbiotic stall latency} (SSL).

This paper quantifies when isolated ceilings fail using \emph{joint speedup synergy} (JSS), the observed joint perfect-branch/perfect-cache speedup divided by the product of the two isolated speedups. Values above one mean independent-ceiling analysis understates attainable gain. Across 53 simulated workloads, 70\% show measurable coupling ($\mathrm{JSS}>1$), although many are close to unity; design-relevant cases concentrate among higher-pressure workloads. With a conservative threshold, 40\% exceed the independence product by more than 6\%, and kernels with $\mathrm{SSO}>20$ show $\mathrm{JSS}$ from 1.23 to 3.29. We introduce \emph{symbiotic stall opportunity} (SSO), a lightweight MPKI-based screen for identifying workloads that merit full joint simulation.

We map high-SSO workloads to four recurring software patterns: neighbor access, hash lookup, linked-structure traversal, and data-dependent modification. We then connect SSL to reorder-buffer occupancy, squash rate, and commit starvation under isolated perfect modes. The result is an evaluation methodology: use SSO to screen, use JSS to validate, and report conditional branch-after-cache and cache-after-branch gains when evaluating branch predictors, prefetchers, caches, or coupled branch/memory mechanisms. Our contribution is a measurement framework showing when isolated perfect modes are adequate and when they understate joint performance headroom.

%% file: introduction.tex
Modern microprocessors rely on aggressive branch predictors and data prefetchers to keep wide out-of-order cores fed \cite{hennessy2011computer}.
Branch mispredictions flush speculative work, while cache misses leave instructions waiting in the reorder buffer (ROB).
Both penalties are severe in irregular applications, where pointer chasing often produces hard-to-predict (H2P) branches whose outcomes depend on recently loaded data \cite{mittal2019survey,bakhshalipour2019evaluation,cavus2021nodetracker}.

Decades of work and dedicated competitions \cite{dpc3,dpc4,cbp5,cbp2025} have advanced branch predictors such as TAGE-SC-L \cite{tage-sc-l,tage} and the Multi-perspective Perceptron \cite{MultiperspectivePP}, as well as prefetchers such as BO \cite{bop}, SPP \cite{spp}, ISB \cite{isb}, VLDP \cite{vldp}, and IMP \cite{imp}.
Other designs explicitly couple control and memory, including address--branch correlation predictors \cite{abc2008}, load-dependent predictors with active updates \cite{exact2010}, and control-flow decoupling \cite{cfd}.
Yet many evaluations still treat branch and cache stalls as separate ceilings: interval and cycles-per-instruction (CPI) analyses attribute cycles to individual miss classes \cite{eyerman2006asplos,eyerman2006ispass,karkhanis2004firstorder,yasin2014topdown}, and architecture studies commonly report perfect-branch \emph{or} perfect-cache speedups as independent upper bounds \cite{karkhanis2004firstorder,lin2019branch,hill2008amdahl}.
This practice implicitly assumes that the joint ceiling is close to the product of the isolated gains.
We test that assumption for irregular hot loops in which a cache-missing load and an H2P branch occur in the same iteration.

Removing one penalty can expose the other.
When cache misses are removed, the loop reaches mispredicted branches more often per unit time, increasing squash pressure.
When branch mispredictions are removed, more correct-path instructions survive, but long-latency loads can fill the ROB and block commit.
Neither isolated perfect mode therefore reveals the joint ceiling.
We call this evaluation pitfall the \emph{fallacy of independent ceilings}.

Triangle counting (\emph{tc}) in GAPBS \cite{gap} illustrates the effect.
Using gem5 \cite{gem5} with TAGE-SC-L and SPP, perfect branch direction alone yields 1.85$\times$ speedup over baseline, and a perfect L1D cache alone yields 1.76$\times$.
If the stalls were independent, jointly perfecting both would be expected near $1.85 \times 1.76 \approx 3.26\times$.
Instead, the combined \texttt{perfect} configuration reaches 5.78$\times$.
The surplus beyond the product is \emph{joint speedup synergy} (JSS): for \emph{tc}, perfecting branch prediction after cache is already improved gives 3.28$\times$ rather than 1.85$\times$, and perfecting cache after branch prediction gives 3.12$\times$ rather than 1.76$\times$.
Each improvement amplifies the marginal value of the other.

Our work does not claim that load--branch coupling is unknown.
Rather, it provides a measurement framework for identifying when isolated evaluation is misleading.
This paper makes the following contributions:
\begin{itemize}
    \item \textbf{Joint speedup synergy (JSS).} We quantify second-order coupling as the ratio of measured joint speedup to the product of isolated perfect-branch and perfect-cache speedups (\Cref{sec:methodology}).
    \item \textbf{Workload screening and taxonomy.} We introduce symbiotic stall opportunity (SSO) as a lightweight screen for workloads that merit joint simulation, and classify high-SSO codes into four recurring software patterns (\Cref{sec:results}).
    \item \textbf{Architectural characterization.} Using gem5 \cite{gem5}, we connect SSL to ROB occupancy, squash frequency, and commit starvation under isolated perfect modes (\Cref{fig:effect_ssl_pipeline,fig:iteration_throughput,fig:upper_bound_ipc}).
    \item \textbf{Evaluation framework.} We provide guidance for reporting conditional branch-after-cache and cache-after-branch gains, selecting high-JSS benchmarks, and interpreting SPEC-style applications versus kernel suites (\Cref{sec:discussion}).
\end{itemize}

The remainder of the paper presents related work (\Cref{sec:related}), motivates SSL (\Cref{sec:motivation}), defines the methodology and metrics (\Cref{sec:methodology}), presents software and hardware results (\Cref{sec:results}), discusses implications (\Cref{sec:discussion}), and concludes.

%% file: related_work.tex
\subsection{Branch prediction and prefetching}
Surveys cover dynamic branch prediction \cite{mittal2019survey,lin2019branch} and hardware data prefetching \cite{bakhshalipour2019evaluation,ayers2020classifying}.
Industrial and academic competitions \cite{cbp5,cbp2025,dpc3,dpc4} have driven state-of-the-art predictors and prefetchers, but usually score the two subsystems separately.
Modern branch predictors are dominated by Perceptron and TAGE-family designs \cite{perceptron-bp,tage,tage-sc-l,MultiperspectivePP}; prefetchers exploit stride, correlation, context, confidence, and linked-data-structure patterns \cite{stride,bop,spp,isb,vldp,imp,karlsson2000prefetching,roth1998dependence,peled2015semantic,braun2019understanding}.
This prior work motivates our baseline mechanisms, but our focus is not a new predictor or prefetcher; it is the evaluation assumption that isolated perfect-mode ceilings compose independently.

\subsection{Coupled control and memory}
Several mechanisms explicitly target branches whose outcomes depend on memory.
The ABC predictor \cite{abc2008} correlates producer load addresses with consumer branch outcomes to resolve long-latency hard-to-predict branches earlier.
EXACT \cite{exact2010} uses load-address context and active store updates to predict dynamic branches.
Node Tracker \cite{cavus2021nodetracker} couples programmable prefetch/pre-execution with branch-outcome streaming on hash-table walks, showing that prefetching alone captures only a small fraction of the available gain until H2P branch mispredictions are also reduced.
Control-flow decoupling \cite{cfd}, slipstream-style dual streams \cite{slipstream}, and runahead or other latency-tolerant execution techniques \cite{slipstream} shorten or bypass dependence chains in amenable loops.
These designs attack the same load--branch coupling that we measure.
Our contribution is complementary: we quantify when isolated upper-bound studies under-report the joint opportunity such mechanisms are meant to exploit.

\subsection{Performance analysis methodology}
Classic CPI stacks and interval-style models decompose execution time into branch, cache, and execution components \cite{eyerman2006asplos,eyerman2006ispass,karkhanis2004firstorder,yasin2014topdown,simulation}.
Interval analysis already recognizes that long data-cache misses can interact with branch penalties by filling the ROB and stalling dispatch \cite{eyerman2006ispass,eyerman2006asplos}.
Amdahl-style reasoning \cite{hill2008amdahl} motivates the multiplicative speedup expectation for independent events.
We make that expectation explicit and test when measured joint perfect-branch/perfect-cache speedup exceeds the product of isolated gains.

Our work is also related to memory-level parallelism (MLP).
Qureshi \emph{et al.} show that isolated cache misses are more performance-critical than overlapped misses and propose MLP-aware replacement \cite{qureshi2006mlp}.
JSS addresses an orthogonal form of overlap: coupling between branch and memory penalties in the same hot loop through ROB occupancy, squash rate, and commit starvation rather than MSHR overlap alone.
The perfect-mode experiments rely on gem5 \cite{gem5}, SimPoint \cite{simpoint}, SMARTS \cite{smarts}, and Lapidary-style checkpointing \cite{lapidary,weisse2019nda,casper2014hardware}.

\subsection{Benchmarks}
Irregular suites such as Olden \cite{olden}, PBBS \cite{pbbs}, AMAC \cite{amac}, GAPBS \cite{gap}, CRONO \cite{crono}, Graph500 \cite{graph500}, and graph libraries \cite{boostGraph,pageRank} stress pointer chasing, graph traversal, hash lookup, and other patterns where load--branch coupling can arise.
SPEC CPU suites \cite{spec} represent full applications; characterization studies show that most SPEC workloads are not simultaneously branch- and memory-bound, with notable exceptions such as MCF \cite{spec}.
Prior graph and memory-system studies \cite{mswp03,imp,amac,cavus2021nodetracker} guide our workload selection, but they do not replace the joint metric developed here: SSO screens for likely coupled pressure, and JSS validates whether isolated perfect-mode ceilings actually understate joint headroom.

%% file: motivation.tex
Hard-to-predict (H2P) branches and irregular memory accesses often meet in the same hot loops of graph and pointer-chasing workloads \cite{olden,gap}.
Prior work proposes coupled predictors and decoupling techniques for this behavior \cite{abc2008,exact2010,cfd}; we ask a complementary evaluation question: what does a conventional wide out-of-order core reveal when only one stall source is removed at a time?
This section defines \emph{symbiotic stall latency} (SSL) and uses Olden \texttt{mst} to show why isolated perfect-branch and perfect-cache studies can understate joint headroom.

\subsection{Symbiotic Stall Latency (SSL)}

\begin{figure}[t]
  \centering
  \resizebox{0.72\columnwidth}{!}{\includegraphics{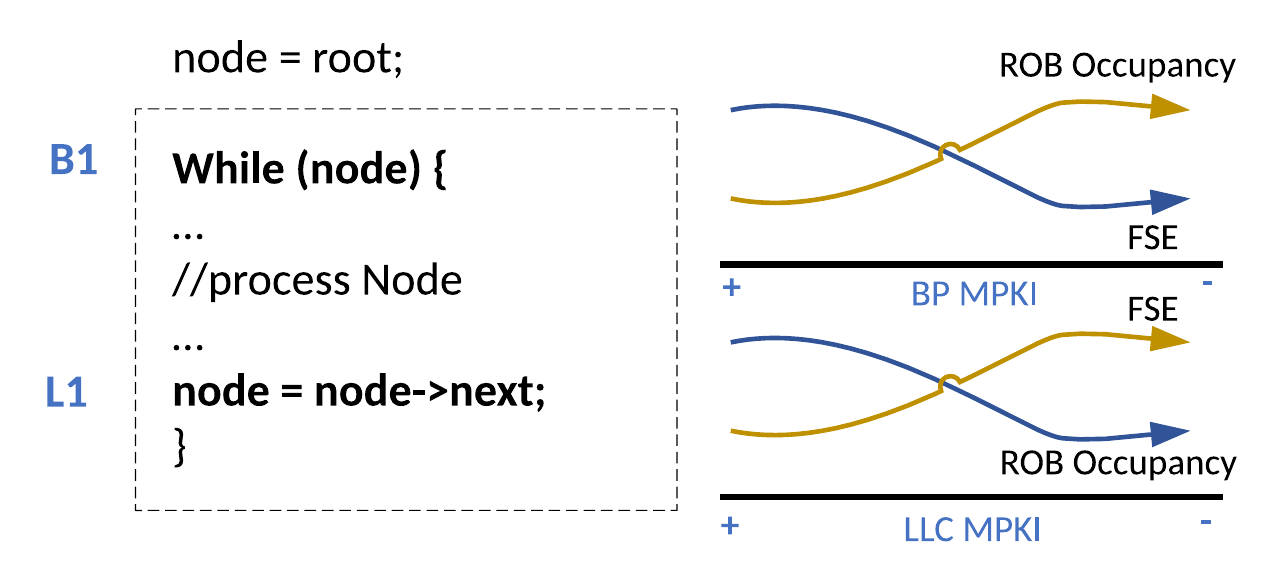}}
  \caption{\textbf{Coupled load--branch stalls in a hot loop.}
  Load \texttt{L1} (irregular access, cache MPKI) pairs with branch \texttt{B1} (data-dependent, branch MPKI).
  \emph{Top path:} better direction prediction lowers branch MPKI but raises ROB occupancy because fewer instructions are squashed while loads still miss.
  \emph{Bottom path:} better caching/prefetching lowers cache MPKI and speeds retirement, but can increase squash events per cycle as the core reaches \texttt{B1} more often.
  \textbf{Insight:} neither isolated curve captures joint headroom; both penalties must be reduced to reach the combined upper bound.}
  \label{fig:arrows}
\end{figure}

Consider a loop where load \verb|L1| often misses in the data cache and dependent branch \verb|B1| is frequently mispredicted.
\Cref{fig:arrows} summarizes the interaction.
If caching or prefetching makes \verb|L1| hit more often, each iteration waits less on memory and reaches the next dynamic \verb|B1| sooner.
The predictor need not mispredict more often per committed instruction, but the machine can see more squash events per cycle because the loop executes faster; throughput then remains limited until branch direction also improves.

The converse also holds.
If direction prediction for \verb|B1| becomes nearly perfect, fewer wrong-path instructions are squashed and more correct-path work survives in the ROB.
That work still depends on misses from \verb|L1|, so the ROB can fill with useful but blocked instructions and stall dispatch, fetch, and commit even though branch MPKI falls.
\textbf{Symbiotic stall latency (SSL)} names this paired behavior: cache-miss and branch-misprediction penalties in the same loop interact through ROB occupancy, squash rate, and commit bandwidth, so removing only one penalty leaves the other as the limiter.

\subsection{Motivating Example: \texttt{mst}}

The \texttt{mst} benchmark computes a minimum spanning tree.
Its hash lookup loop is a concrete instance of \Cref{fig:arrows}: branch \texttt{B1} and pointer-chase load \texttt{L1} (\texttt{ent=ent->next}) depend on linked-list data (\Cref{tab:mst_code}).

\begin{table}[!ht]
\centering
\caption{Hash lookup loop in \texttt{mst} (C, left; x86, right).}
\label{tab:mst_code}
\definecolor{lightGrey}{rgb}{0.9, 0.9, 0.9}
\newcommand{\Hilight}{\makebox[0pt][l]{\color{lightGrey}\rule[-0.45em]{0.1\linewidth}{1.5em}}}
\vspace{2pt}
\codepairstart
\begin{lstlisting}[
  style=codeside,
  escapechar=\%,
  language={c},
  morekeywords={B1,B2,B3,B4,B5,B6,B7,L1},
  label={mst_c},
]
void *HashLookup(unsigned int key, Hash hash)
{
    ...

B1  for(ent = hash->array[j]; ent && ent->key != key;
L1      ent=ent->next);
    if(ent) return ent->entry;
    return NULL;
}
\end{lstlisting}
\codepairmid
\begin{lstlisting}[
  style=codeside,
  morekeywords={B1,L1,C1},
  label={mst_assembly},
]
    ...
    mov     0x0(%rbp), %rdx
    cltq
    mov     (%rdx, %rdx, 8), %rdx
    test    $rax, %rax
    jne     0x401e89    <...>
    ...
L1  mov     0x10(%rax), %rax
    ...
    cmp     %ebx, (%rax)
B1  jne     0x401e80    <...>

\end{lstlisting}
\codepairend
\end{table}

\begin{figure}[!t]
  \centering
  \begin{subfigure}{\columnwidth}
    \centering
    \includegraphics[width=0.92\columnwidth]{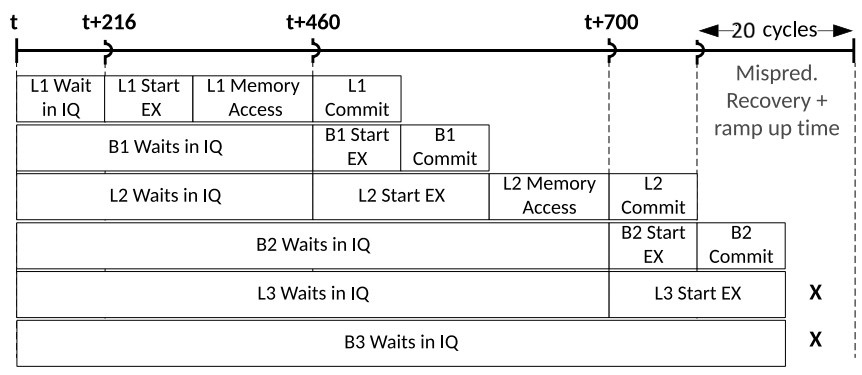}
    \caption{\texttt{base}: both \texttt{L1} misses and \texttt{B1} mispredictions limit progress.}
    \label{fig:ssl_pipeline_base}
  \end{subfigure}
  \begin{subfigure}{\columnwidth}
    \centering
    \includegraphics[width=0.92\columnwidth]{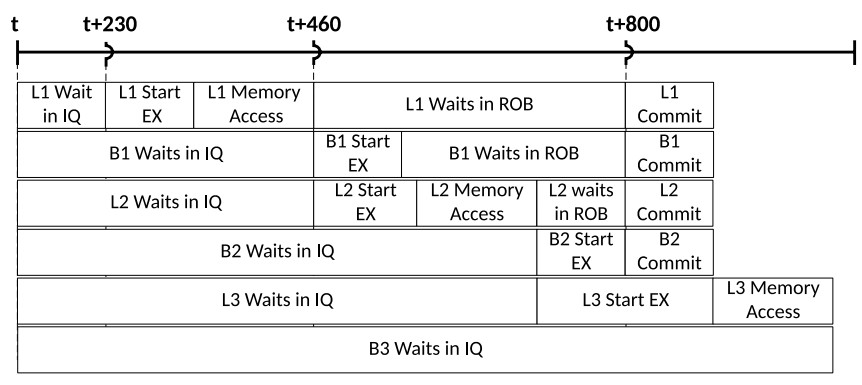}
    \caption{\texttt{perfect-bp}: fewer squashes, but ROB fills on surviving load latency.}
    \label{fig:ssl_pipeline_pbp}
  \end{subfigure}
  \begin{subfigure}{\columnwidth}
    \centering
    \includegraphics[width=0.92\columnwidth]{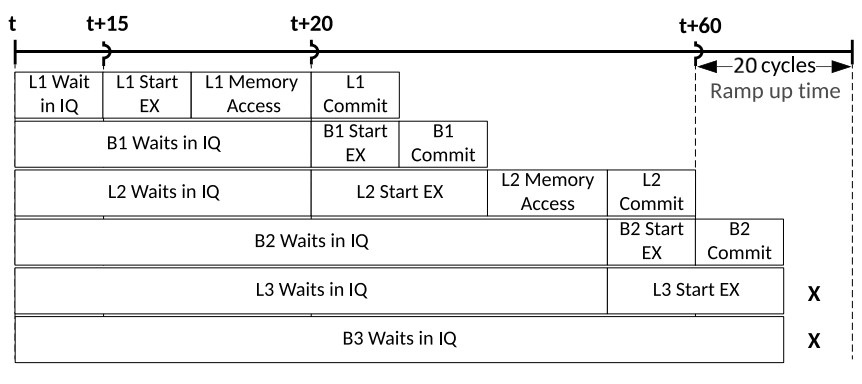}
    \caption{\texttt{perfect-cache}: faster memory, but frequent squashes on \texttt{B1}.}
    \label{fig:ssl_pipeline_pcache}
  \end{subfigure}
  \begin{subfigure}{\columnwidth}
    \centering
    \includegraphics[width=0.92\columnwidth]{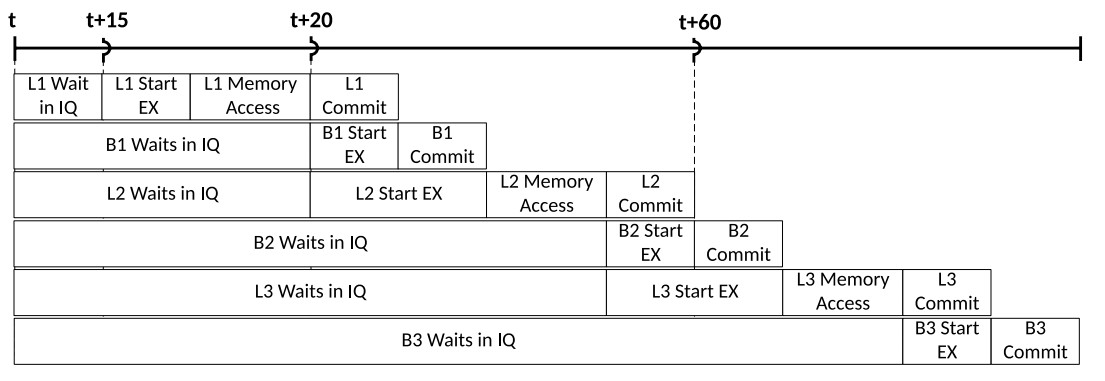}
    \caption{\texttt{perfect}: short drain; both penalties removed in the hot loop.}
    \label{fig:ssl_pipeline_perf}
  \end{subfigure}
  \caption{\textbf{Pipeline snapshots of SSL on the \texttt{mst} hash loop.}
  Three iterations of the branch--load pair are shown; time scales differ across panels.
  The isolated perfect modes still leave either surviving load latency or wrong-path churn, whereas joint \texttt{perfect} yields the short, commit-ready loop body that JSS quantifies in \Cref{sec:methodology}.}
  \label{fig:ssl_pipeline}
\end{figure}

We simulate four gem5 configurations (\Cref{sec:methodology}): \texttt{baseline} (TAGE-SC-L plus the best prefetcher per benchmark), \texttt{perfect-bp} (recorded correct branch direction), \texttt{perfect-cache} (all demand accesses hit in the L1D timing path while preserving modeled L1 latency), and \texttt{perfect} (both).
\Cref{fig:ssl_pipeline} illustrates the resulting ROB behavior.
With both penalties active, \texttt{B1} mispredictions flush speculative chains while \texttt{L1} misses leave dependents waiting.
With \texttt{perfect-bp}, squashes drop, but surviving pointer-chase iterations fill the ROB behind long-latency loads.
With \texttt{perfect-cache}, the loop reaches \texttt{B1} more often per unit time, so wrong-path work dominates despite faster memory.
Only \texttt{perfect} shortens both the load wait and branch recovery, producing the compact, useful in-flight loop body needed to approach the joint ceiling.

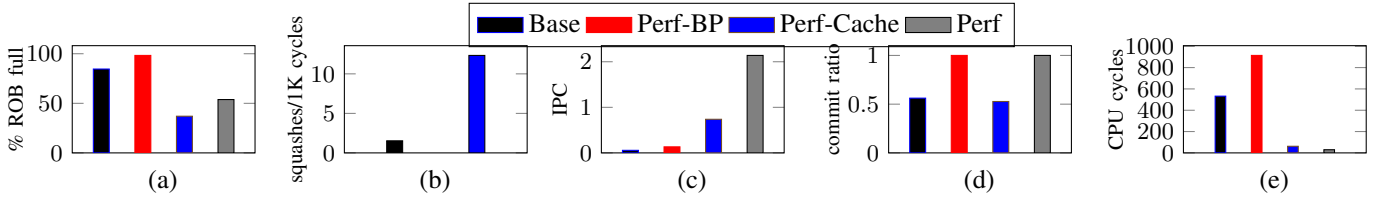
\begin{figure*}[!t]
\begin{tikzpicture}

 \node[below, node distance=0cm, yshift=-0.1cm, xshift = 1.2cm] {(a)};
 \node[below, node distance=0cm, yshift=-0.1cm, xshift = 4.8cm] {(b)};
 \node[below, node distance=0cm, yshift=-0.1cm, xshift = 8.2cm] {(c)};
 \node[below, node distance=0cm, yshift=-0.1cm, xshift = 12cm] {(d)};
 \node[below, node distance=0cm, yshift=-0.1cm, xshift = 15.9cm] {(e)};
\begin{axis}[
    ybar,
    ylabel={\% ROB full},
    y label style={at={(0.2,0.5)}, font=\footnotesize},
    tick label style={font=\small},
    width=4cm,
    height=3cm,
    xtick=\empty,
    ymin=0,
    bar width=6,
    enlarge x limits=1,
    legend entries={Base, Perf-BP,Perf-Cache, Perf},
        legend style={at={(1.0\columnwidth,1.4)}, anchor=north, legend columns=-1},
    y tick label style={/pgf/number format/.cd,%
          scaled y ticks = false,
          set thousands separator={},
          fixed
    },
                legend image code/.code={
        \draw [#1] (0cm,-0.1cm) rectangle (0.2cm,0.2cm);
    },
    ]

    \addplot +[area legend, fill=black] coordinates {(1,84.57914094)};
    \addplot  +[area legend, fill=red]coordinates {(2,98.30831073) };
    \addplot  +[area legend, fill=blue]coordinates {(3,37.00485419) };
    \addplot  +[area legend, fill=gray]coordinates {(4,53.79182584) };
\end{axis}
\begin{axis}[
    ybar,
    width=4cm,
    height=3cm,
    ylabel={squashes/1K cycles},
    y label style={at={(+0.25,0.5)}, font=\footnotesize},
    tick label style={font=\small},
    xshift=3.6cm,
    bar width=6,
    xtick=\empty,
    ymin=0,
    enlarge x limits=1,
    legend entries={},
    y tick label style={/pgf/number format/.cd,%
          scaled y ticks = false,
          set thousands separator={},
          fixed
    },
    ]

    \addplot +[draw=black, fill=black] coordinates {(1,  1.527494353)};
    \addplot  +[draw=black,, fill=blue]coordinates {(3,12.33377706) };
\end{axis}
\begin{axis}[
    ybar,
    width=4cm,
    height=3cm,
    ylabel={IPC},
    y label style={at={(+0.28,0.5)}, font=\footnotesize},
    tick label style={font=\small},
    xshift=7cm,
    bar width=6,
    xtick=\empty,
    ymin=0,
    enlarge x limits=1,
    legend entries={},
    y tick label style={/pgf/number format/.cd,%
          scaled y ticks = false,
          set thousands separator={},
          fixed
    },
    ]

    \addplot +[area legend, fill=black] coordinates {(1,  0.056583)};
    \addplot  +[area legend, fill=red]coordinates {(2,0.1327839714) };
    \addplot  +[area legend, fill=blue]coordinates {(3,0.7377677498) };
    \addplot  +[area legend, fill=gray]coordinates {(4,2.138978066) };
\end{axis}
\begin{axis}[
    ybar,
    width=4cm,
    height=3cm,
    bar width=6,
    ylabel={commit ratio},
    y label style={at={(+0.2,0.5)}, font=\footnotesize},
    tick label style={font=\small},
    xshift=10.8cm,
    xtick=\empty,
    ymin=0,
    enlarge x limits=1,
    legend entries={},
    y tick label style={/pgf/number format/.cd,%
          scaled y ticks = false,
          set thousands separator={},
          fixed
    },
    ]

    \addplot +[area legend, fill=black] coordinates {(1,  0.5617812785)};
    \addplot  +[area legend, fill=red]coordinates {(2,1) };
    \addplot  +[area legend, fill=blue]coordinates {(3,0.5273963909) };
    \addplot  +[area legend, fill=gray]coordinates {(4,1) };
\end{axis}
\begin{axis}[
    ybar,
    width=4cm,
    height=3cm,
    bar width=4,
    ylabel={CPU cycles},
    y label style={at={(+0.15,0.5)}, font=\footnotesize},
    tick label style={font=\small},
    xshift=14.7cm,
    xtick=\empty,
    ymin=0,
    enlarge x limits=1,
    legend entries={},
    y tick label style={/pgf/number format/.cd,%
          scaled y ticks = false,
          set thousands separator={},
          fixed
    },
    ]

    \addplot +[area legend, fill=black] coordinates {(1,  532.349)};
    \addplot  +[area legend, fill=red]coordinates {(2,914.333) };
    \addplot  +[area legend, fill=blue]coordinates {(3,62.2195) };
    \addplot  +[area legend, fill=gray]coordinates {(4,29.7475) };
\end{axis}
\end{tikzpicture}
  \caption{\textbf{\texttt{mst} hash-loop behavior under four simulation configurations.}
  Bars compare \texttt{baseline}, \texttt{perfect-bp}, \texttt{perfect-cache}, and \texttt{perfect}.
  (a)~ROB occupancy rises with \texttt{perfect-bp} because fewer squashes leave long-latency loads in flight.
  (b)~Squashes per 1K cycles jump under \texttt{perfect-cache} when memory is fast but \texttt{B1} remains mispredicted.
  (c)~IPC shows isolated modes far below \texttt{perfect}.
  (d)~Iteration commit ratio: fraction of observed loop iterations that eventually commit.
  (e)~Snapshot drain cycles.
  \textbf{Observation:} \texttt{perfect-bp} commits all tracked iterations but pays high cycle cost; \texttt{perfect-cache} drains quickly but wastes half the iterations to squashes; only \texttt{perfect} combines full commit ratio and short drain time, yielding JSS of 1.23$\times$ beyond the isolated-speedup product (2.35$\times$ $\times$ 13.04$\times$ vs.\ 37.82$\times$ measured).}
  \label{fig:mst_ssl}
\end{figure*}

If \texttt{L1} and \texttt{B1} were independent bottlenecks, the speedup of \texttt{perfect} would be close to the product of isolated \texttt{perfect-bp} and \texttt{perfect-cache} speedups.
The quantitative metrics in \Cref{fig:mst_ssl} show otherwise.

\paragraph{Iteration tracking.}
We tag one instruction per hot-loop iteration and periodically snapshot ROB contents until every tagged instruction commits or squashes.
The \emph{iteration commit ratio} (\Cref{fig:mst_ssl}d) is committed iterations divided by observed iterations in the snapshot.
Under \texttt{perfect-bp}, nearly all observed iterations commit, but snapshots drain more slowly because loads still miss; IPC rises only to 2.35$\times$ baseline.
Under \texttt{perfect-cache}, snapshots drain quickly, but only ${\sim}$53\% of iterations commit because squashes rise roughly 8$\times$.
Joint \texttt{perfect} achieves commit ratio $=1$ with the lowest drain time, producing 37.82$\times$ baseline IPC.

\paragraph{Joint speedup synergy (JSS) for \texttt{mst}.}
The isolated speedups are 2.35$\times$ (\texttt{perfect-bp}) and 13.04$\times$ (\texttt{perfect-cache}), whose product is 30.6$\times$.
The measured joint speedup is 37.82$\times$, so JSS is $37.82/(2.35 \times 13.04) \approx 1.23$: joint perfecting exceeds the independence baseline by 23\%.
This surplus is not a numerical artifact; it matches a visible ROB-state change in which joint perfecting increases useful iteration commit ratio and reduces snapshot drain time.
The rest of the paper generalizes this measurement across suites, defines Symbiotic Stall Opportunity (SSO) as a cheaper screener, and catalogs software patterns with similar coupling.

%% file: methodology.tex
\subsection{Workloads}

The motivating example in \Cref{sec:motivation} shows that SSL arises when an irregular load and an H2P branch repeatedly meet in the same hot loop.
We therefore study suites that stress irregular memory and H2P branches: Olden~\cite{olden}, PBBS~\cite{pbbs}, AMAC~\cite{amac}, GAPBS~\cite{gap}, CRONO~\cite{crono}, Graph500~\cite{graph500}, and PageRank on BGL~\cite{boostGraph,pageRank}.
Olden and Graph500 emphasize pointer-based traversals; PBBS and GAPBS include STL- and template-heavy graph kernels; AMAC and CRONO provide hash-, array-, and graph-oriented codes.
Together, they expose kernels where load--branch coupling is plausible but not guaranteed.

\textbf{GAP vs.\ CRONO.}
We include both GAPBS and CRONO because suite names alone are insufficient for SSL studies.
Both target graph analytics, but their implementation styles differ: GAPBS uses C++ kernels with pointer-heavy neighbor access (e.g., \texttt{tc}), whereas CRONO uses C arrays with more direct indexing and multi-threaded drivers \cite{gap,crono}.
On our metrics, GAP \texttt{tc} and several Olden/PBBS kernels show high SSO and $\mathrm{JSS} \gg 1$, whereas most CRONO workloads (\texttt{crono\_bc}, \texttt{crono-bfs}, \texttt{Crono\_sssp}, \texttt{Crono\_tc}) cluster near $\mathrm{JSS} \approx 1$; only \texttt{Crono\_cc} clears the SSO screen.
Thus, implementation style can determine whether the independence fallacy appears, even within graph benchmarks.
SPEC CPU2017 \cite{spec} is evaluated with the same framework and discussed in \Cref{sec:discussion}, where phase behavior becomes the main issue.

\subsection{Simulation Methodology}

All simulations use gem5 \cite{gem5} in syscall-emulation mode for x86-64.
Because SE-mode execution is deterministic, every configuration is reproducible cycle-for-cycle; reported IPC, speedup, and JSS values therefore carry no run-to-run variance, and differences across modes reflect microarchitectural behavior rather than sampling noise.
\Cref{tab:CPU-config} lists the modeled out-of-order core, which follows contemporary high-performance processor parameters \cite{hennessy2011computer,parkhurst2006single}.
The configuration is intentionally conventional: the goal is to isolate how branch and memory idealizations compose on a wide out-of-order baseline, not to propose a new core.
All workloads are compiled with GCC 14.3 using \texttt{-O2 -g -static -march=x86-64}; kernel inputs follow standard suite configurations and exercise the hot traversal, hash-probe, or graph-processing regions studied in \Cref{sec:results}.

    \begin{table}[htb]
    \scriptsize
    \caption{Simulated microarchitecture parameters.}
    \label{tab:CPU-config}
    \begin{center}
    \begin{tabular}{ | l | l | } 
    \hline
    \textbf{Core} & \shortstack[l]{Out-of-order superscalar,\\Fetch/decode/issue/commit width: 8\\256-entry ROB; 96-entry LQ; 64-entry SQ} \\
    \hline
    \textbf{Branch predictor} & \shortstack[l]{TAGE-SC-L (8\,KB total) \cite{tage,tage-sc-l}\\20-cycle branch misprediction latency\\4096-entry BTB; 16-entry RAS\\512-entry indirect target predictor} \\
    \hline
    \textbf{Level-1 data (L1D) cache} & 64\,KB, 4-cycle, 8-way, 24 MSHRs \\
    \hline
    \textbf{Level-2 (L2) cache} & 256\,KB, 12-cycle, 8-way, 24 MSHRs \\
    \hline
    \textbf{Last-level cache (LLC)} & 4\,MB, 32-cycle, 16-way, 48 MSHRs \\
    \hline
    \textbf{Prefetchers evaluated} & Stride, SPP \cite{spp}, BOP \cite{bop}, IMP \cite{imp} \\
    \hline
    \textbf{Reported baseline} & Best prefetcher per benchmark \\
    \hline
    \end{tabular}
    \end{center}
    \end{table}

\paragraph{Regions of interest.}
Kernel-style benchmarks use one checkpoint immediately before the hot traversal region, followed by 5M warmup instructions and 100M detailed instructions.
This focuses the kernel study on regions where SSL can occur rather than setup or teardown code.
SPEC CPU2017 integer-speed workloads use SMARTS-style sampling \cite{smarts} with native checkpoints captured using Lapidary-style tooling \cite{lapidary,casper2014hardware}.
Each SPEC checkpoint warms for 5M instructions and then runs 1M detailed instructions; we simulate 100 checkpoints per benchmark, which prior work finds sufficient for accurate whole-program SPEC coverage \cite{weisse2019nda}.
Averaged statistics summarize whole-program behavior, and selected checkpoints support the phase analysis in \Cref{sec:discussion}.

\paragraph{Simulation configurations.}
We compare four modes throughout the paper.
\texttt{baseline} uses TAGE-SC-L with the best evaluated prefetcher for each benchmark.
\texttt{perfect-bp} replaces branch-direction predictions with recorded correct taken/not-taken outcomes while keeping the same memory system.
\texttt{perfect-cache} makes all demand data accesses hit in the L1D timing path while retaining TAGE-SC-L and the benchmark's best prefetcher.
\texttt{perfect} applies both idealizations.
These modes are diagnostic ceilings, not predictions for a deployable predictor or prefetcher: they test whether improving one subsystem changes the marginal value of improving the other.
Absolute JSS may vary with core width, predictor, cache hierarchy, memory latency, and threading; we treat it as a composability check.
We study single-threaded SE-mode execution; OS, multicore, and shared-cache effects are left as future work.

\subsection{Measurement Methodology and Metrics}
We report performance in IPC.

\subsubsection{Perfect Branch Prediction}
Perfect branch prediction emulates \emph{direction} only: taken vs.\ not-taken outcomes are replayed from a trace recorded in a first-pass run over committed branches.
A second pass overrides the TAGE-SC-L decision with the recorded bitstream.
Indirect branches and BTB mismatches can still squash, but these events are rare across our kernel suite ($>$99.9\% accuracy).
On squash, a ROB-sized circular buffer rewinds the trace pointer so recovery remains consistent with in-order commit.

\subsubsection{Perfect Cache}
\texttt{perfect-cache} removes L1D miss latency while preserving the modeled 4-cycle L1 access in timing mode.
Accesses below L1 use gem5 functional mode with zero added latency, approximating an infinitely large L1D with instant refill from backing storage.
This is an upper bound on cache-side benefit, not a realistic hierarchy.
Address translation still completes; we do not model a separate perfect DTLB, so remaining TLB costs appear in all configurations.
Unless noted otherwise, MPKI statistics count demand accesses and exclude prefetch accesses when computing SSO.

\subsubsection{Instruction attribution}
For detailed case studies in \Cref{sec:results}, we attribute static PCs that contribute more than 10\% of branch mispredictions or L1/L2 demand misses in the hot loop.
PCs that contribute to both event classes identify the load--branch slices discussed in \Cref{sec:motivation} and connect aggregate SSO/JSS values to the software patterns analyzed next.

\subsubsection{Symbiotic Stall Opportunity (SSO)}
SSO is an inexpensive screener applied before running all four perfect configurations.
It asks whether a workload has enough committed branch-misprediction and demand data-cache-miss pressure for SSL to be plausible:
\begin{equation} \label{eq:sso}
        \mathrm{SSO} = \frac{\mathrm{MPKI}_{BP} \times \mathrm{MPKI}_{CACHE}}{\mathrm{MPKI}_{BP} + \mathrm{MPKI}_{CACHE}}
\end{equation}
SSO is not a stall-cycle estimate; it is a screening proxy for simultaneous branch and data-cache pressure.
Here, $\mathrm{MPKI}_{BP}$ counts committed branch mispredictions per thousand instructions, excluding wrong-path branches, and $\mathrm{MPKI}_{CACHE}$ counts demand data-cache misses, excluding prefetch accesses.
The harmonic-product form is high only when both event classes are frequent, so SSO identifies workloads that merit perfect-mode JSS measurement.

\begin{figure*}[!t]
\centering
\begin{subfigure}[t]{\textwidth}
\centering
\includegraphics[width=\textwidth]{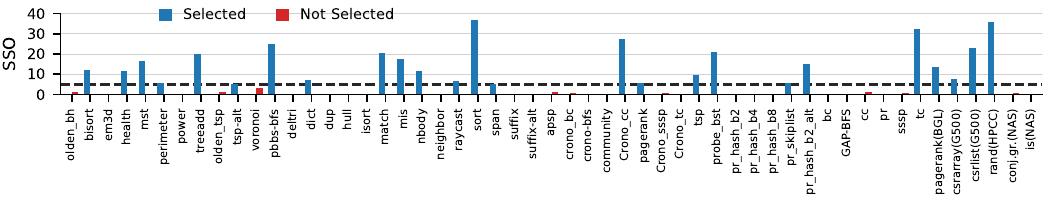}
\caption{\textbf{SSO screening.}
Blue bars: benchmarks with SSO $>$ 5 selected for detailed study; red bars: not selected.
Dashed line: SSO $= 5$ cutoff (\Cref{eq:sso}).
Low branch/cache MPKI kernels rarely show second-order effects.}
\label{fig:sso_screening}
\end{subfigure}
\begin{subfigure}[t]{\textwidth}
\centering
\includegraphics[width=\textwidth]{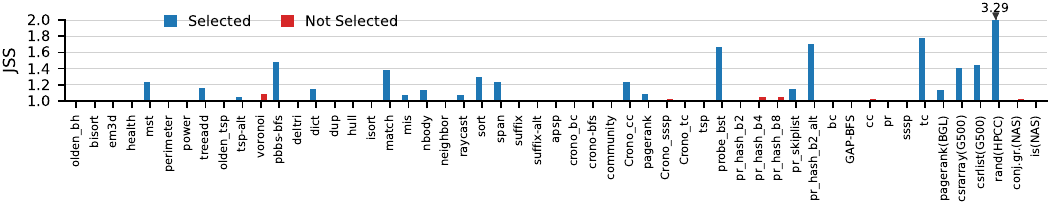}
\caption{\textbf{Measured JSS validation.}
Same kernel order and blue/red encoding as (\subref{fig:sso_screening}).
High-SSO kernels usually exceed the baseline; \texttt{bisort} and \texttt{health} are exceptions with elevated SSO but $\mathrm{JSS}\approx 1$.}
\label{fig:sso_jss_validate}
\end{subfigure}
\caption{\textbf{Population-wide SSO screening and JSS measurement for all simulated kernels.}
Inexpensive MPKI-based screening (\Cref{eq:sso}) is followed by perfect-mode JSS validation (\Cref{eq:jss}).
\textbf{Observation:} high SSO is a useful screen for $\mathrm{JSS}>1$, but it must be validated with perfect-mode simulation.}
\label{fig:sso_jss}
\end{figure*}

\subsubsection{Joint Speedup Synergy (JSS)}
JSS validates whether isolated perfect-mode ceilings compose.
Following Amdahl-style independence reasoning \cite{hill2008amdahl}, if branch and cache improvements act on separate bottlenecks, the joint speedup should equal the product of isolated gains.
We use that product as the implicit composability assumption behind isolated perfect-mode ceilings, and measure how far the observed joint mode departs from it:
\begin{equation} \label{eq:expectedSpeedup}
        \mathrm{Expected}_{\mathrm{speedup}} = \mathrm{Gain}_{pBP} \times \mathrm{Gain}_{pCache}
\end{equation}
\begin{equation}\label{eq:gainpBP}
        \mathrm{Gain}_{pBP} = {IPC_{pBP}}/{IPC_{\mathrm{base}}}
\end{equation}
\begin{equation}\label{eq:gainpCache}
        \mathrm{Gain}_{pCache} = {IPC_{pCache}}/{IPC_{\mathrm{base}}}
\end{equation}

\pgfplotstableread[row sep=\\,col sep=&,format=inline]{
Benchmark &	Selected &	Not_Selected \\
olden_bh &  0 & 1.1 \\
bisort	&	12.3	&		\\
em3d	&	0	&	0.2	\\
health	&	11.7	&	0	\\
mst	&	16.5	&	0	\\
perimeter	&	5.7	&	0	\\
power	&	0	&	0	\\
treeadd	&	20.2	&	0	\\
olden_tsp	&	0	&	1.2	\\
tsp-alt	&	5.4	&	0	\\
voronoi	&	0	&	3.2	\\
pbbs-bfs	&	25.1	&	0	\\
deltri	&	0	&	0	\\
dict	&	7.2	&	0	\\
dup	&	0	&	0	\\
hull	&	0	&	0	\\
isort	&	0	&	0	\\
match	&	20.6	&	0	\\
mis	&	17.4	&	0	\\
nbody	&	11.5	&	0	\\
neighbor	&	0	&	0	\\
raycast	&	6.6	&	0	\\
sort	&	36.9	&	0	\\
span	&	5.3	&	0	\\
suffix	&	0	&	0	\\
suffix-alt	&	0	&	0	\\
apsp	&	0	&	1.2	\\
crono_bc	&	0	&	0.7	\\
crono-bfs	&	0	&	0.2	\\
community	&	0	&	0.1	\\
Crono_cc	&	27.3	&	0	\\
pagerank	&	5.6	&	0	\\
Crono_sssp	&	0	&	0.9	\\
Crono_tc	&	0	&	0	\\
tsp	&	9.5	&	0	\\
probe_bst	&	21	&	0	\\
pr_hash_b2	&	0	&	0	\\
pr_hash_b4	&	0	&	0	\\
pr_hash_b8	&	0	&	0	\\
pr_skiplist	&	5.7	&	0	\\
pr_hash_b2_alt	&	15.2	&	0	\\
bc	&	0	&	0	\\
GAP-BFS	&	0	&	0	\\
cc	&	0	&	1.1	\\
pr	&	0	&	0	\\
sssp	&	0	&	0.9	\\
tc	&	32.4	&	0	\\
pagerank(BGL)	&	13.7	&	0	\\
csrarray(G500)	&	7.8	&	0	\\
csrlist(G500)	&	23	&	0	\\
rand(HPCC)	&	35.9	&	0	\\
conj.gr.(NAS)	&	0	&	0.8	\\
is(NAS)	&	0	&	0	\\
    }\loadbranch

\pgfplotstableread[row sep=\\,col sep=&,format=inline]{
Benchmark &	Selected &	Not_Selected \\
olden_bh 	&	0	&	0.9799470466	\\
bisort	&	0.9324895724	&	0	\\
em3d	&	0	&	1.006118404	\\
health	&	0.8673142318	&	0	\\
mst	&	1.235453323	&	0	\\
perimeter	&	1.013505006	&	0	\\
power	&	0	&	0.9998010677	\\
treeadd	&	1.164062971	&	0	\\
olden_tsp	&	0	&	0.9899619299	\\
tsp-alt	&	1.054540286	&	0	\\
voronoi	&	0	&	1.080579881	\\
pbbs-bfs	&	1.48578988	&	0	\\
deltri	&	0	&	1	\\
dict	&	1.149916373	&	0	\\
dup	&	0	&	1	\\
hull	&	0	&	1.002776748	\\
isort	&	0	&	1.000093986	\\
match	&	1.383682193	&	0	\\
mis	&	1.073064029	&	0	\\
nbody	&	1.134671734	&	0	\\
neighbor	&	0	&	1.001109528	\\
raycast	&	1.07613757	&	0	\\
sort	&	1.290216255	&	0	\\
span	&	1.235618802	&	0	\\
suffix	&	0	&	1.000000096	\\
suffix-alt	&	0	&	1.000000795	\\
apsp	&	0	&	1.001562913	\\
crono_bc	&	0	&	1.009190316	\\
crono-bfs	&	0	&	1.007487726	\\
community	&	0	&	0.9997171805	\\
Crono_cc	&	1.230306061	&	0	\\
pagerank	&	1.085001664	&	0	\\
Crono_sssp	&	0	&	1.023424735	\\
Crono_tc	&	0	&	1.000176164	\\
tsp	&	0.999150431	&	0	\\
probe_bst	&	1.663839618	&	0	\\
pr_hash_b2	&	0	&	1.000038707	\\
pr_hash_b4	&	0	&	1.050710976	\\
pr_hash_b8	&	0	&	1.050710976	\\
pr_skiplist	&	1.144609724	&	0	\\
pr_hash_b2_alt	&	1.701345203	&	0	\\
bc	&	0	&	0.999872282	\\
GAP-BFS	&	0	&	1.000000002	\\
cc	&	0	&	1.024798396	\\
pr	&	0	&	1.00014811	\\
sssp	&	0	&	0.968745634	\\
tc	&	1.772536008	&	0	\\
pagerank(BGL)	&	1.131445894	&	0	\\
csrarray(G500)	&	1.404937757	&	0	\\
csrlist(G500)	&	1.438867137	&	0	\\
rand(HPCC)	&	3.288297461	&	0	\\
conj.gr.(NAS)	&	0	&	1.023146578	\\
is(NAS)	&	0	&	1.000000001	\\
    }\overlaplatency

The observed joint speedup is
\begin{equation} \label{eq:speedupObserved}
        \mathrm{Speedup}_{\mathrm{observed}} = {IPC_{\mathrm{perfect}}}/{IPC_{\mathrm{base}}}
\end{equation}
and JSS captures surplus beyond the independence baseline:
\begin{equation} \label{eq:jss}
        \mathrm{JSS} = {\mathrm{Speedup}_{\mathrm{observed}}}/{\mathrm{Expected}_{\mathrm{speedup}}}
\end{equation}
Values above 1 indicate that jointly perfecting branch and cache exceeds the product of isolated perfect modes; values near 1 indicate that the isolated ceilings compose well enough for first-order CPI-style analysis.

\Cref{tab:conditional_speedup} makes this test concrete.
For representative high-SSO kernels, columns~4--5 report the gain from applying one perfect mode after the other.
When a conditional gain exceeds its isolated counterpart in columns~2--3, improving one subsystem has increased the marginal value of improving the other, so the isolated ceilings are not composable.

\begin{table}[!t]
\centering
\scriptsize
\caption{Conditional speedups on high-SSO kernels ($\mathrm{Gain} = \mathrm{IPC}_{\mathrm{mode}}/\mathrm{IPC}_{\mathrm{base}}$). Columns~2--3 are isolated perfect modes; columns~4--5 apply the second perfect mode after the first. Column~6 is JSS.}
\label{tab:conditional_speedup}
\begin{tabular}{lccccc}
\hline
\textbf{Kernel} &
\multicolumn{2}{c}{\textbf{Isolated}} &
\multicolumn{2}{c}{\textbf{After other ceiling}} &
\textbf{JSS} \\
\cline{2-5}
& \texttt{p-bp} & \texttt{p-cache} & \texttt{p-bp}$^\dagger$ & \texttt{p-cache}$^\ddagger$ & \\
\hline
\texttt{tc}       & 1.85 & 1.76 & 3.27 & 3.12 & 1.77 \\
\texttt{match}       & 1.24 & 3.22 & 1.72 & 4.45 & 1.38 \\
\texttt{sort}       & 1.59 & 1.69 & 2.05 & 2.19 & 1.30 \\
\texttt{probe-bst}& 1.27 & 1.30 & 2.11 & 2.16 & 1.66 \\
\texttt{bfs}      & 1.12 & 4.04 & 1.66 & 6.01 & 1.49 \\
\texttt{cc}     & 1.80 & 1.64 & 2.22 & 2.02 & 1.23 \\
\texttt{rand} & 1.01 & 3.61 & 3.32 & 11.88 & 3.29 \\
\hline
\end{tabular}
\parbox{\columnwidth}{\footnotesize $^\dagger$Measured with \texttt{perfect-cache} already active; $^\ddagger$with \texttt{perfect-bp} already active.}
\end{table}

\subsection{Benchmark Selection}
\Cref{fig:sso_jss} closes the methodology loop by applying SSO and then validating with measured JSS for every simulated kernel.
We use SSO $> 5$ as the screening threshold for detailed analysis in \Cref{sec:results}, but exclude \texttt{bisort} and \texttt{health}.
Both clear the SSO threshold yet show JSS $\leq 1$ because \texttt{perfect-cache} partially substitutes for branch prediction improvement rather than exposing it.
In these benchmarks, the dominant mispredicting branch resolves on a value loaded by the dominant cache-missing load: the load is on the branch's input dependency chain rather than competing with it for ROB resources.
When \texttt{perfect-cache} removes the load latency, the branch-resolving data arrives faster, shortening recovery even without predictor improvement.
This differs from SSL, where load and branch penalties compete through ROB occupancy and squash rate.
SSO cannot distinguish these cases, so high-SSO workloads require JSS validation before motivating coupled branch/memory mechanisms.

Across all 53 simulated kernels in \Cref{fig:sso_jss_validate}, the JSS distribution has three regimes:
\begin{itemize}[nosep,leftmargin=*]
\item \textbf{Near-independent behavior} ($\mathrm{JSS}\approx 1$): isolated perfect-branch and perfect-cache ceilings compose well enough for first-order analysis.
\item \textbf{Measurable coupling} ($\mathrm{JSS}>1$): 37 kernels (70\%) exceed the independence product, although many are close to unity.
\item \textbf{Actionable headroom} ($\mathrm{JSS}>1.06$ or high-SSO/high-JSS): 21 kernels (40\%) exceed the product by more than 6\%, and kernels with $\mathrm{SSO}>20$ show $\mathrm{JSS}$ from 1.23 to 3.29.
\end{itemize}
Thus, low SSO or near-unity JSS supports conventional isolated analysis, while high SSO plus elevated JSS identifies workloads where branch and memory mechanisms should be evaluated jointly.

The next section explains these high-JSS cases at the software level, by grouping hot loops into recurring load--branch patterns, and at the hardware level, by relating those patterns to ROB occupancy, squash rate, commit starvation, and speedup under the four simulation modes.

%% file: analysis.tex
This section explains the source of the high-JSS cases selected in \Cref{sec:methodology}.
The goal is to connect the metric result to concrete program structure and pipeline behavior: first, which hot-loop patterns repeatedly couple cache-missing loads with H2P branches, and second, why removing only one stall source changes the bottleneck instead of eliminating it.

\subsection{Software-Level Analysis}
\input{software}

\subsection{Hardware-Level Analysis}
The software taxonomy identifies where SSL originates; we now examine how it appears in the out-of-order engine.
Using the same screened kernels and four simulation modes, we measure ROB occupancy, squash rate, idle commit cycles, loop-level iteration throughput, ROB-size sensitivity, and normalized speedup (\Cref{fig:effect_ssl_pipeline,fig:iteration_throughput,fig:rob_effect,fig:upper_bound_ipc}).
Together, these measurements show why the joint \texttt{perfect} mode can exceed the product of isolated perfect modes.
\input{Hardware}

%% file: software.tex
The methodology in \Cref{sec:methodology} identifies high-JSS kernels; here we explain what those kernels share.
JSS is not tied to benchmark names, but to hot-loop structures that pair irregular loads with data-dependent branches.
We group screened kernels into four descriptive families (\Cref{tab:category}) derived from static-PC attribution and inspection of dominant hot loops, not from suite labels.
The taxonomy explains recurring sources of high JSS; it is not a formal partition and does not claim that every irregular workload belongs uniquely to one category.
For representative members, we report hot loops, dominant PCs, and measured $\mathrm{JSS}$ (\Cref{eq:jss}).

\begin{table}[!t]
    \scriptsize
    \caption{Algorithmic families with high symbiotic stall opportunity.}
    \label{tab:category}
    \centering
    \begin{tabular}{ | l | l | } 
    \hline
    \textbf{Family} & \textbf{Representative benchmarks} \\
    \hline
    Neighboring node access (NNA) & BFS, CC, Match, MIS, PageRank, TC \\
    \hline
    Hash table lookup (HTL) & BST, Hash, Skiplist, MST, Dict, RandAcc \\
    \hline
    Linked-structure traversal (LDST) & Treeadd, TSP, Health, Perimeter \\
    \hline
    Data-dependent modification (DDM) & Sort, CSR-List, Bisort, CSR-Array \\
    \hline
    \end{tabular}
\end{table}
    
\subsubsection{Neighboring Node Access (NNA)}
NNA kernels traverse adjacency lists or edge-relaxation streams.
The loaded neighbor ID determines both the next access and the branch that gates traversal, so cache-missing loads feed hard-to-predict compare branches.
These kernels show high JSS because the pair repeats at graph-edge granularity.

\begin{table}[!ht]
\centering
\caption{GAP \texttt{tc} triangle-counting hot loop (C, left; x86, right).}
\label{gap-tc-snip}
\codepairstart
\begin{lstlisting}[
  style=codeside,
  escapechar=\%,
  language={c},
  morekeywords={B1,B2,C1,C2,L1,L2},
]
for (NodeID u = 0; u < g.num_nodes(); u++) {
L1    for (NodeID v : g.out_neigh(u)) {
B1      if (v > u) break;
          auto it = g.out_neigh(u).begin();
L2      for (NodeID w : g.out_neigh(v)) {
B2        if (w > v) break;
            while (*it < w) it++;
            if (w == *it) total++;
}}}
\end{lstlisting}
\codepairmid
\begin{lstlisting}[
  style=codeside,
  morekeywords={B1,B2,C1,C2,L1,L2},
]
    cmp     %rbp,%r12 
    je      0x41bd88
    mov     %rbp,%r10
    ... 
L1  mov    (%r10),%edi
    cmp    %ebx,%edi 
B1  jg     0x41bd88
    movslq %edi,%rax
    add    $0x1,%rax  
    ...
L2  mov    (%rsi),%ecx 
    cmp    %ecx,%edi 
B2  jl     0x41bd7f 
\end{lstlisting}
\codepairend
\end{table}

GAP \texttt{tc} \cite{gap} is the canonical NNA example (\Cref{gap-tc-snip}).
\texttt{OrderedCount} nests neighbor loops: loads \texttt{L1}/\texttt{L2} walk adjacency lists, while branches \texttt{B1}/\texttt{B2} break on vertex ordering.
Their inputs are loaded neighbor IDs, so prediction quality and cache behavior are coupled.
In 100M detailed instructions, \texttt{B1}/\texttt{B2} account for 44\%/27\% of mispredictions and \texttt{L1}/\texttt{L2} for 39\%/28\% of demand misses.
The resulting $\mathrm{JSS}=1.77$ shows that either isolated fix understates joint headroom.

\subsubsection{Hash Table Lookups (HTL)}
HTL kernels probe chains, trees, or FIFO structures keyed by a hash or search value.
Each probe combines an irregular pointer load with an equality, ordering, or traversal branch.
Olden-style tables scan linked buckets until a key matches, as in \texttt{mst} (\Cref{sec:motivation}); AMAC-style probes \cite{amac} keep in-flight keys in FIFO slots embedded in structures such as a binary-search tree.
In both styles, unpredictable probe length creates H2P branches while pointer chasing creates demand misses, so one dynamic iteration supplies both sides of JSS.
\begin{table}[!ht]
\centering
\caption{C and x86 assembly of BST (C, left; x86, right).}
\label{amac-bst-snip}
\definecolor{lightGrey}{rgb}{0.9, 0.9, 0.9}
\newcommand{\Hilight}{\makebox[0pt][l]{\color{lightGrey}\rule[-0.45em]{0.1\linewidth}{1.5em}}}
\codepairstartasym{0.63}{0.32}
\begin{lstlisting}[
  style=codeside,
  escapechar=\%,
  language={c},
  morekeywords={B1,C1,L1},
]
while (i < rel->num_tuples) {
    k = (k >= SIZE) ? 0 : k;
    if (fifo[k].ptr) {
        comp = fifo[k].ptr->tuple.key - fifo[k].key;
B1      if (comp == 0) {
            outR->tuples[fifo[k].payload].key = 
L1                fifo[k].ptr->tuple.key;
            ...
B1      } else if (comp > 0) {
            fifo[k].ptr = fifo[k].ptr->right;}
        ...
    } else {
L1      fifo[k].key = rel->tuples[i].key;
        fifo[k].ptr = bt->nodes;
    }
}
\end{lstlisting}
\codepairmid
\begin{lstlisting}[
  style=codeside,
  morekeywords={B1,C1,L1},
]
    shl   $0x4,%rcx
    add   (%r14),%rcx
L1  mov   %r10,(%rcx)
    mov   %rsi,0x8(%rcx)
    ...
    add   (%r12),%rcx
L1  mov   (%rcx),%rcx
    mov   %rsi,0x10(%rax)
    ...
    jae   0x4029c8
    cmp   %edi,%ebx
B1  jg    0x402938
    mov   $0x1,%edi
    ...
B1  jle   0x4029f0 
     ... 
\end{lstlisting}
\codepairend
\end{table}
AMAC \texttt{probe-bst} illustrates the second style (\Cref{amac-bst-snip}).
Its cascaded \texttt{B1} compares implement the three-way decision on \texttt{comp} and account for 95\% of mispredictions; paired \texttt{L1} loads account for 99\% of misses.
The kernel reaches $\mathrm{JSS}=1.66$, so the dominant branch and load PCs are mutually limiting.

\subsubsection{Linked Data Structure Traversal (LDST)}
LDST kernels walk trees, lists, or perimeter structures with little work per node.
Null checks or leaf tests supply H2P branches, and child-pointer loads supply misses.
Their JSS is more mixed than NNA or HTL because traversal alone is insufficient: the penalties must repeatedly overlap.

\begin{table}[!h]
\centering
\caption{C and x86 assembly of Treeadd (C, left; x86, right).}
\label{olden-treeadd-snip}
\definecolor{lightGrey}{rgb}{0.9, 0.9, 0.9}
\newcommand{\Hilight}{\makebox[0pt][l]{\color{lightGrey}\rule[-0.45em]{0.1\linewidth}{1.5em}}}
\codepairstartasym{0.60}{0.35}
\begin{lstlisting}[
  style=codeside,
  escapechar=\%,
  language={c},
  morekeywords={B1,L1,L2},
]
    int TreeAdd (tree_t *t)  {
B1  if (t == NULL)  return 0;
    else {
    ...
    tree_t *tleft, *tright;
L2  tleft = t->left;            
    leftval = TreeAdd(tleft); 
    tright = t->right;       
L1  rightval = TreeAdd(tright); 
    value = t->val;             
    return leftval + rightval + value;}}
\end{lstlisting}
\codepairmid
\begin{lstlisting}[
  style=codeside,
  morekeywords={B1,B2,L2,L1},
]
    ...
B1  je     0x401dc0 <...>
    ...
L2  mov    0x8(%rdi),%rdi 
    callq  0x401d90 <...>
L1  mov    0x10(%rbp),%rdi 
    mov    %eax,%ebx
    callq  0x401d90 <...>
    ...
    retq   
    ...
\end{lstlisting}
\codepairend
\end{table}
Olden \texttt{treeadd} (\Cref{olden-treeadd-snip}) is representative.
The \texttt{B1} null check accounts for 95\% of mispredictions, while child-pointer loads \texttt{L1}/\texttt{L2} account for 99\% of misses.
Its $\mathrm{JSS}=1.16$ is above independence but lower than the strongest NNA/HTL cases, motivating PC attribution and JSS validation.

\subsubsection{Data Dependent Modifications (DDM)}
DDM kernels modify arrays or graph structures while scanning, partitioning, or reordering data.
Their branches depend on loaded values, and their memory references are often scattered or bidirectional.
Here coupling need not be pointer chasing; data-dependent comparisons and in-place updates can expose misses and branch recovery in the same loop.
        \begin{table}[h]
\centering
\caption{C and x86 assembly of Comparison Sort (C, left; x86, right).}
\label{pbbs-sort-snip}
\definecolor{lightGrey}{rgb}{0.9, 0.9, 0.9}
\newcommand{\Hilight}{\makebox[0pt][l]{\color{lightGrey}\rule[-0.45em]{0.1\linewidth}{1.5em}}}
\codepairstart
\begin{lstlisting}[
  style=codeside,
  escapechar=\%,
  language={c},
  morekeywords={B1,B2,L1,L2B2,L1},
]
        while (true) {
B1      while (comp(first, pivot))
        ++first;
        --last;
L2B2    while (comp(pivot, last))
        --last;
        if (!(first < last))    return first;
        std::iter_swap(first, last);
L1      ++first;  }
\end{lstlisting}
\codepairmid
\begin{lstlisting}[
  style=codeside,
  morekeywords={B1,B2,L1,L2,L1},
]
    ...
B1  ja     0x41d936 
    ...
L2  movsd  -0x8(%rax),%xmm1 
    sub    $0x8,%rax
    comisd %xmm0,%xmm1
B2  ja     0x41d910 
    ...
L1  movsd  (%rdx),%xmm2 
   ...
\end{lstlisting}
\codepairend
\end{table}
    
PBBS comparison sort (\Cref{pbbs-sort-snip}) uses IntroSort partitioning.
Branches \texttt{B1}/\texttt{B2} contribute 77\% of mispredictions, while loads \texttt{L1}/\texttt{L2} contribute 86\% of misses.
Its $\mathrm{JSS}=1.29$ shows that data-dependent partitioning creates the same pitfall as graph traversal or hash lookup.

\subsubsection{Cross-family JSS summary}
\begin{figure*}[!t]
\centering
\includegraphics[width=\textwidth]{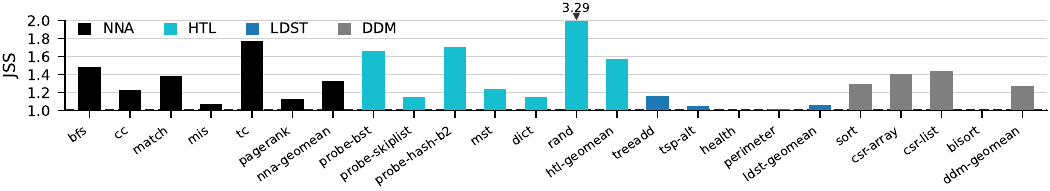}
\caption{\textbf{Joint speedup synergy (JSS) by benchmark and algorithmic family.}
Each bar is $\mathrm{JSS}=\mathrm{Speedup}_{\mathrm{observed}}/\mathrm{Expected}_{\mathrm{speedup}}$ (\Cref{eq:jss}); color encodes family (\Cref{tab:category}).
Family membership alone does not determine JSS: NNA/HTL concentrate high surplus, while several LDST kernels stay near 1.0.}
\label{fig:sw-benchmark}
\end{figure*}

\Cref{fig:sw-benchmark} reorganizes the screened kernels by software family rather than suite.
Membership in an irregular algorithmic family is not sufficient for strong coupling: high JSS requires branch and memory penalties to be frequent, co-located in the same dynamic region, and mutually limiting.
Geometric-mean surplus beyond independence is approximately 57\% for HTL, 33\% for NNA, 27\% for DDM, and 6\% for LDST.
HTL and NNA repeatedly pair irregular loads with data-dependent branches at probe or edge granularity.
DDM shows strong coupling when scattered access coincides with unpredictable partitioning decisions, while LDST is closer to independence because many traversals do not sustain enough simultaneous branch and memory pressure.
Within a family, implementation style matters: pointer-heavy neighbor traversal can yield high JSS, whereas direct array indexing of a similar algorithm can remain near independence.
High SSO likewise does not guarantee high JSS: \texttt{bisort} and \texttt{health} are counterexamples in which faster load completion partially accelerates branch resolution, making the effects more substitutive than mutually limiting.
The next subsection connects these patterns to ROB pressure, squash pressure, and idle commit cycles.

\pgfplotstableread[row sep=\\,col sep=&,format=inline]{
Benchmark & NNA & HTL & LDST & DDM\\
bfs & 1.48578988 & 0 & 0 & 0\\
cc & 1.230306061 & 0 & 0 &0 \\
match & 1.383682193 & 0 & 0 & 0\\
mis & 1.073064029 & 0 & 0 &0 \\
tc & 1.772536008 & 0 & 0 &0 \\
pagerank & 1.131445894 & 0 & 0 & 0\\
nna-geomean & 1.326305491 & 0 & 0 &0 \\
probe-bst & 0 & 1.663839618 & 0 &0 \\
probe-skiplist & 0 & 1.144609724 & 0 &0 \\
probe-hash-b2 & 0 & 1.701345203 & 0 &0 \\
mst & 0 & 1.235453323 & 0 & 0\\
dict & 0 & 1.149916373 & 0 & 0\\
rand & 0 & 3.288297461 & 0 &0 \\
htl-geomean & 0 & 1.572790359 & 0 &0 \\
treeadd & 0 & 0 & 1.164062971 & 0\\
tsp-alt & 0 & 0 & 1.054540286 & 0\\
health & 0 & 0 & 1.004540286 & 0\\
perimeter & 0 & 0 & 1.013505006 & 0\\
ldst-geomean & 0 & 0 & 1.057324332 & 0 \\
sort & 0 & 0 & 0 & 1.290216255\\
csr-array  & 0 & 0 & 0 & 1.404937757\\
csr-list & 0 & 0 & 0 & 1.438867137 \\
bisort & 0 & 0 & 0 & 1.000093986\\
ddm-geomean & 0 & 0 & 0 & 1.270852875\\
    }\mehmet

%% file: Hardware.tex
The software families in \Cref{fig:sw-benchmark} identify where SSL originates; the hardware metrics show how those loops stress the out-of-order core.
The perfect modes are upper-bound experiments: \texttt{perfect-bp} idealizes committed branch direction while still allowing BTB and indirect-target effects, \texttt{perfect-cache} preserves L1D timing while making every demand access hit at L1 latency, and \texttt{perfect} applies both.

\begin{figure}[!t]
\begingroup
\captionsetup[subfigure]{skip=1pt,aboveskip=0pt,belowskip=0pt}
\vspace{-2pt}
\centering
    \begin{subfigure}[t]{\columnwidth}
 \begin{tikzpicture}
    \begin{axis}[
    ybar,
    ylabel={Avg.\ occupied ROB entries},
    y label style={at={(0.02,0.5)}, font=\footnotesize},
    every x tick label/.append style={font=\scriptsize},
    width=\columnwidth,
    height=3.75cm,
    x tick label style={rotate=75},
    symbolic x coords={
    Bisort,CC,Health,MST,Pagerank,bst,Hash,skiplist,RandAcc,CSR-List,Comp.Sort,TC,Treeadd,TSP,geomean,
            },
    ymin=0,
    ymax=280,
    xtick=data,
    bar width=3,
    enlarge x limits=0.05,
    legend entries={base, perf-BP},
        legend style={at={(0.5,1.02)}, anchor=south, legend columns=-1, font=\scriptsize},
    y tick label style={/pgf/number format/.cd,%
          scaled y ticks = false,
          set thousands separator={},
          fixed
    },
        legend image code/.code={
        \draw [#1] (0cm,-0.1cm) rectangle (0.2cm,0.2cm);
    },
    ]
    
   \addplot+ table[x=Benchmark,y=base]{\robOccup};
    \addplot+ table[x=Benchmark,y=perfect-bp]{\robOccup};
\end{axis}
\end{tikzpicture}
       \subcaption{\textbf{Time-averaged occupied ROB entries (256-entry ROB, \Cref{tab:CPU-config}).}
       Bars report occupied slots per cycle, not percent of capacity.
       \texttt{perfect-bp} retains more in-flight work when squashes drop.}
       \label{fig:rob_occupancy}
    \end{subfigure}
     \begin{subfigure}[t]{\columnwidth}
 \begin{tikzpicture}
    \begin{axis}[
    ybar,
    ylabel={Squash events/1K cycles},
    y label style={at={(0.05,0.5)}, font=\footnotesize},
    every x tick label/.append style={font=\scriptsize},
    width=\columnwidth,
    height=3.75cm,
    x tick label style={rotate=75}, 
    symbolic x coords={
      Bisort,CC,Health,MST,Pagerank,bst,Hash,skiplist,RandAcc,CSR-List,Comp.Sort,TC,Treeadd,TSP,geomean,
            },
    ymin=0,
    xtick=data,
    bar width=3,
    enlarge x limits=0.05,
    legend entries={base, perf-cache},
        legend style={at={(0.5,1.02)}, anchor=south, legend columns=-1, font=\scriptsize},
    y tick label style={/pgf/number format/.cd,%
          scaled y ticks = false,
          set thousands separator={},
          fixed
    },
            legend image code/.code={
        \draw [#1] (0cm,-0.1cm) rectangle (0.2cm,0.2cm);
    },
    ]
    
   \addplot table[x=Benchmark,y=base]{\squashevents};
    \addplot table[x=Benchmark,y=perfect-cache]{\squashevents};
\end{axis}
\end{tikzpicture}
       \subcaption{\textbf{Squash events per 1K cycles.}
       \texttt{perfect-cache} can \emph{increase} squashes by reaching data-dependent branches faster.}
       \label{fig:squash_events_per_cycle}
    \end{subfigure}
     \begin{subfigure}[t]{\columnwidth}
 \begin{tikzpicture}
    \begin{axis}[
    ybar,
    ylabel={Idle commit cycles (\%)},
    y label style={at={(0.05,0.5)}, font=\footnotesize},
    every x tick label/.append style={font=\scriptsize},
    width=\columnwidth,
    height=3.75cm,
    x tick label style={rotate=75}, 
    symbolic x coords={
      Bisort,CC,Health,MST,Pagerank,bst,Hash,skiplist,RandAcc,CSR-List,Comp.Sort,TC,Treeadd,TSP,geomean,
            },
    ymin=0,
    xtick=data,
    bar width=1,
    enlarge x limits=0.04,
    legend entries={base , perf-bp , perf-cache , perf},
        legend style={at={(0.5,1.02)}, anchor=south, legend columns=-1, font=\scriptsize},
    y tick label style={/pgf/number format/.cd,%
          scaled y ticks = false,
          set thousands separator={},
          fixed
    },
            legend image code/.code={
        \draw [#1] (0cm,-0.1cm) rectangle (0.2cm,0.2cm);
    },
    ]
    
   \addplot table[x=Benchmark,y=base]{\commitcycles};
    \addplot table[x=Benchmark,y=perfect-bp]{\commitcycles};
    \addplot table[x=Benchmark,y=perfect-cache]{\commitcycles};
    \addplot table[x=Benchmark,y=perfect]{\commitcycles};
\end{axis}
\end{tikzpicture}
       \subcaption{\textbf{Fraction of cycles with no commit.}
       Isolated perfect modes still leave many idle commits; \texttt{perfect} sharply reduces them.}
       \label{fig:commit_cycles_ratio}
    \end{subfigure}
\caption{\textbf{Pipeline stress under isolated versus joint perfect modes.}
(a)~Fewer squashes expose memory-bound ROB heads under \texttt{perfect-bp}.
(b)~Faster loops mispredict more often per unit time under \texttt{perfect-cache}.
(c)~Joint \texttt{perfect} cuts idle commits on coupled-stall kernels (geomean ${\sim}$76\%$\rightarrow$24\%).
Together, the panels show why $\mathrm{JSS}>1$: one perfect source shifts, rather than removes, the bottleneck.}
\label{fig:effect_ssl_pipeline}
\vspace{-2pt}
\endgroup
\end{figure}
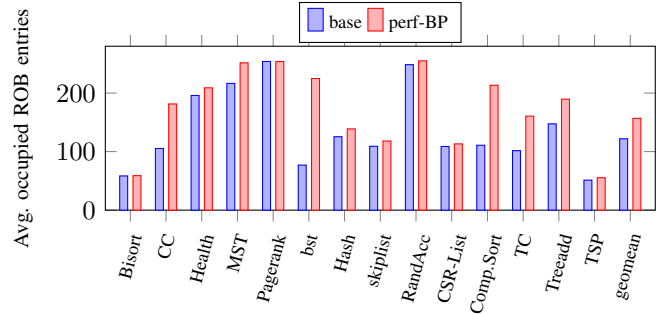
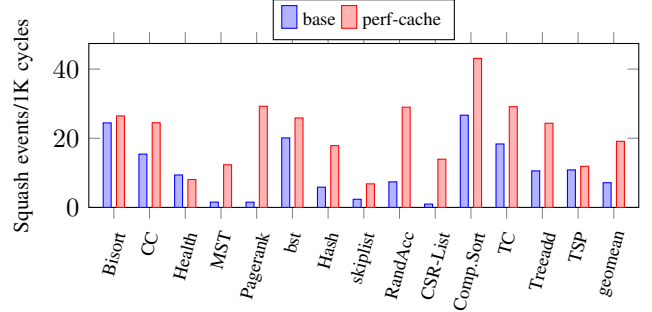
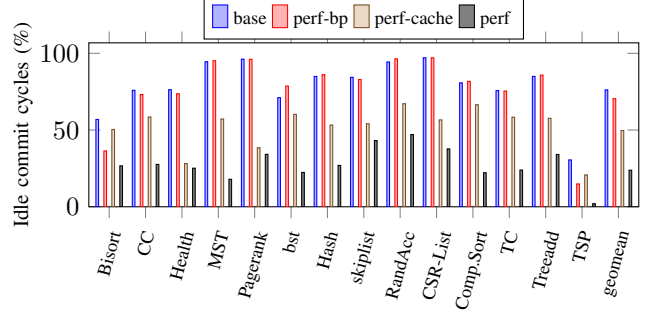

\Cref{fig:effect_ssl_pipeline} reports three pipeline signatures on screened kernels.
Under \texttt{perfect-bp}, fewer squashes leave more correct-path instructions in flight; if they depend on irregular loads, the ROB fills with useful but blocked work (geomean occupied entries ${\sim}$122$\rightarrow$157).
Under \texttt{perfect-cache}, faster loads can make the core reach the same H2P branch more often per unit time, increasing squashes per 1K cycles (geomean ${\sim}$7$\rightarrow$19).
Cycles with no committed instruction remain high under isolated modes but drop from ${\sim}$76\% in \texttt{base} to ${\sim}$24\% under joint \texttt{perfect}.
Thus, perfecting one latency source often converts the other into the limiter instead of removing SSL.

\pgfplotstableread[row sep=\\,col sep=&,format=inline]{
Benchmark &	base &	perfect-bp\\
Bisort &	58.47675591 &	58.9436208\\
CC &	105.3063227 &	181.4345671\\
Health &	195.8987246 &	208.9852291\\
MST &	216.5226008 &	251.6692755\\
Pagerank &	253.9839127 &	253.8054218\\
bst &	76.9587219 &	224.8367768\\
Hash &	125.4478462 &	138.7667577\\
skiplist &	109.0840564 &	118.0880849\\
RandAcc &	248.5354942 &	255.0127177\\
CSR-List &	108.7408561 &	113.2144419\\
Comp.Sort &	110.8649157 &	213.4379461\\
TC &	101.6632739 &	160.6206017\\
Treeadd &	147.4193762 &	189.6048545\\
TSP&	51.25143526 &	55.38865811\\
geomean &	121.9443006 &	156.8754361\\
    }\robOccup
\pgfplotstableread[row sep=\\,col sep=&,format=inline]{
Benchmark &	base &	perfect-cache\\
Bisort &	24.43719555 &	26.46514477\\
CC &	15.40989467 &	24.47941025\\
Health &	9.378465574 &	8.044603526\\
MST &	1.527494353 &	12.33377706\\
Pagerank &	1.516127115 &	29.25890279\\
bst &	20.10295556 &	25.87043857\\
Hash &	5.85E+00 &	17.85204363\\
skiplist &	2.355311383 &	6.810620485\\
RandAcc &	7.394922395 &	28.98947804\\
CSR-List &	0.9634177934 &	13.9359579\\
Comp.Sort &	26.66743777 &	43.09164923\\
TC &	18.34774643 &	29.11748527\\
Treeadd &	10.56360439 &	24.32999723\\
TSP &	10.84284216 &	11.90238722\\
geomean &	7.145484492 &	19.11334174\\
    }\squashevents
\pgfplotstableread[row sep=\\,col sep=&,format=inline]{
Benchmark & base & perfect-bp & perfect-cache & perfect\\
Bisort & 56.9113224 & 36.3739992 & 50.35514664 & 26.66200437\\
CC & 75.89229093 & 73.11090905 & 58.49766974 & 27.61238831\\
Health & 76.27272981 & 73.63563528 & 28.15648526 & 25.15150804\\
MST & 94.49090679 & 95.13671794 & 57.19706234 & 17.92558826\\
Pagerank & 96.17157587 & 96.1001399 & 38.45116869 & 34.16502927\\
bst & 71.10352389 & 78.63021107 & 60.22433357 & 22.39122638\\
Hash & 84.91522701 & 86.05472746 & 53.24236183 & 26.94465435\\
skiplist & 84.35524204 & 82.91653334 & 54.1180062 & 43.11807795\\
RandAcc & 94.33653564 & 96.39105234 & 67.15775284 & 47.04798301\\
CSR-List & 97.09538662 & 97.04884877 & 56.62262848 & 37.70148181\\
Comp.Sort & 80.74483264 & 81.71878344 & 66.47837015 & 22.17460168\\
TC & 75.66477155 & 75.40933824 & 58.34867625 & 23.91697653\\
Treeadd & 84.94928024 & 85.78537382 & 57.64158675 & 34.07974804\\
TSP & 30.53272232 & 14.90594321 & 20.74809888 & 2.006551471\\
geomean & 76.17035066 & 70.4401709 & 49.68314436 & 23.82411203\\
    }\commitcycles

\subsubsection{Iteration throughput in hot loops}
Pipeline counters show core pressure; loop tracking shows whether it is useful work or wrong-path churn.
For representative loops, we tag one static instruction per iteration, snapshot the ROB, and track each tag until commit or squash.
The \emph{iteration commit ratio} is committed tags divided by observed tags; \emph{snapshot drain time} is cycles to drain the snapshot, normalized to \texttt{base}.

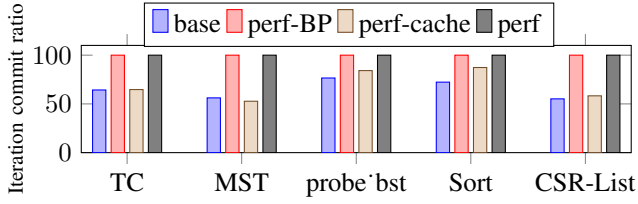
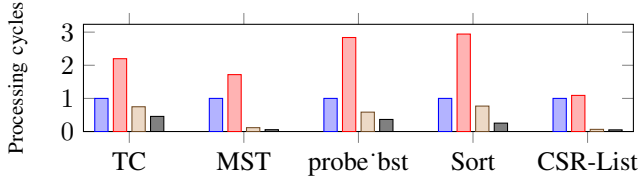
\begin{figure}[htbp]
\centering
    \begin{subfigure}[b]{1\columnwidth}
 \begin{tikzpicture}
    \begin{axis}[
    ybar,
    ylabel={Iteration commit ratio},
    y label style={at={(0.05,0.5)},  font=\footnotesize},
    every y label/.append style={font=\scriptsize},
    width=\columnwidth,
    height=3cm,
    x tick label style={rotate=0}, 
    symbolic x coords={
    TC,MST,probe_bst,Sort,CSR-List
    },
    ymin=0,
    xtick=data,
    bar width=5,
    enlarge x limits=0.1,
    legend entries={base, perf-BP, perf-cache, perf},
        legend style={at={(0.4\columnwidth,1.4)}, anchor=north, legend columns=-1},
    y tick label style={/pgf/number format/.cd,%
          scaled y ticks = false,
          set thousands separator={},
          fixed
    },
        legend image code/.code={
        \draw [#1] (0cm,-0.1cm) rectangle (0.2cm,0.2cm);
    },
    ]
    
   \addplot+ table[x=Benchmark,y=base]{\iterationRatio};
    \addplot+ table[x=Benchmark,y=perfect-bp]{\iterationRatio};
    \addplot+ table[x=Benchmark,y=perfect-cache]{\iterationRatio};
    \addplot+ table[x=Benchmark,y=perfect]{\iterationRatio};
\end{axis}
\end{tikzpicture}
       \subcaption{\textbf{Hot-loop iteration commit ratio.}
       \texttt{perfect-cache} on \texttt{mst}/\texttt{csr-list} wastes roughly half of observed iterations to squashes.}
       \label{figIt:iteration_commit_ratio}
    \end{subfigure}
    \\
     \begin{subfigure}[b]{1\columnwidth}
 \begin{tikzpicture}
    \begin{axis}[
    ybar,
    ylabel={Processing cycles},
    y label style={at={(0.05,0.5)},  font=\footnotesize},
    width=\columnwidth,
    height=3cm,
    x tick label style={rotate=0}, 
    symbolic x coords={
    TC,MST,probe_bst,Sort,CSR-List
    },
    ymin=0,
    xtick=data,
    bar width=5,
    enlarge x limits=0.1,
    legend entries={},
        legend style={at={(0.4\columnwidth,1.2)}, anchor=north, legend columns=-1},
    y tick label style={/pgf/number format/.cd,%
          scaled y ticks = false,
          set thousands separator={},
          fixed
    },
    ]
    
   \addplot+ table[x=Benchmark,y=base]{\snapshotTime};
    \addplot+ table[x=Benchmark,y=perfect-bp]{\snapshotTime};
    \addplot+ table[x=Benchmark,y=perfect-cache]{\snapshotTime};
    \addplot+ table[x=Benchmark,y=perfect]{\snapshotTime};
\end{axis}
\end{tikzpicture}
       \subcaption{\textbf{Snapshot drain time (normalized to \texttt{base}).}
       Values below 1.0 mean faster loop completion; \texttt{perfect-bp} can \emph{slow} drain when memory dominates.}
       \label{figIt:processing_time}
    \end{subfigure}

\caption{\textbf{Hot-loop iteration throughput (five representative kernels).}
\texttt{perfect-bp} recovers speculative iterations but can lengthen drain time on memory-bound loops; \texttt{perfect-cache} shortens snapshots but not when squashes dominate.
Joint \texttt{perfect} approaches the short, fully committed loop body in \Cref{fig:mst_ssl}.}
\label{fig:iteration_throughput}

\end{figure}

\Cref{fig:iteration_throughput} shows the \texttt{mst} trade-off across five kernels.
With \texttt{perfect-bp}, nearly all observed iterations commit, but drain time can rise because surviving work still waits on loads.
With \texttt{perfect-cache}, snapshots drain quickly, but \texttt{mst} and \texttt{csr-list} commit only ${\sim}$50--60\% of observed iterations because squashes waste work.
Only joint \texttt{perfect} combines high commit ratio with short drain time.

\pgfplotstableread[row sep=\\,col sep=&,format=inline]{
Benchmark & base & perfect-bp & perfect-cache & perfect\\
TC & 64.28577168 & 100 & 64.70575317 & 100\\
MST & 56.17812785 & 100 & 52.73963909 & 100\\
probe_bst & 76.57145417 & 100 & 84.16663523 & 100\\
Sort & 72.32463785 & 100 & 87.32401627 & 100\\
CSR-List & 55.16454364 & 100 & 58.28693454 & 100\\
    }\iterationRatio
\pgfplotstableread[row sep=\\,col sep=&,format=inline]{
Benchmark & base & perfect-bp & perfect-cache & perfect\\
TC & 1 & 2.198006717 & 0.7479679672 & 0.4566884093\\
MST & 1 & 1.717544318 & 0.1168772741 & 0.05587969546\\
probe_bst & 1 & 2.837281141 & 0.5851308186 & 0.3646704474\\
Sort & 1 & 2.941560645 & 0.7679988569 & 0.2534250368\\
CSR-List & 1 & 1.090430047 & 0.06549126393 & 0.05043541147\\
    }\snapshotTime

\subsubsection{Effect of ROB size}
We next test whether SSL is simply a finite-window artifact.
We scale the ROB from 256 to 1024 entries, with IQ, LQ/SQ, and physical registers scaled proportionally (\Cref{tab:CPU-config}), and re-run \texttt{perfect-bp} and \texttt{perfect-cache} on four high-occupancy kernels.

\pgfplotstableread[row sep=\\,col sep=&,format=inline]{
Benchmark & 256entries & 512entries & 1024entries\\
bfs & 0.24 & 0.30 & 0.33\\
bisort & 1.42 & 1.51 & 1.56\\
cc & 0.90 & 1.23 & 1.71\\
pagerank & 0.83 & 0.99 & 1.17\\
    }\robPerfBp

\pgfplotstableread[row sep=\\,col sep=&,format=inline]{
Benchmark & 256entries & 512entries & 1024entries\\
bfs & 0.867 & 0.868 & 0.869\\
bisort & 1.270 & 1.253 & 1.264\\
cc & 0.820 & 0.820 & 0.820\\
pagerank & 1.741 & 1.726 & 1.703\\
}\robPerfCache

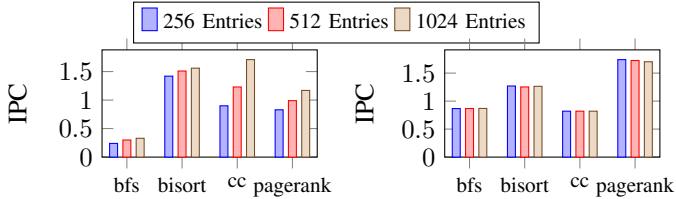
\begin{figure}[!t]
    \centering
    \begin{subfigure}[t]{0.5\columnwidth}
        \centering
        \begin{tikzpicture}
    \begin{axis}[
    ybar,
    ylabel={IPC},
    y label style={at={(0.04,0.5)}},
    every x tick label/.append style={font=\footnotesize},
    width=\linewidth,
    height=3cm,
    x tick label style={rotate=0}, 
    symbolic x coords={
    bfs,bisort,cc,pagerank    },
    ymin=0,
    xtick=data,
    bar width=3,
    enlarge x limits=0.15,
    legend entries={256 Entries, 512 Entries, 1024 Entries},
        legend style={at={(0.7\columnwidth,1.45)}, fill=white, font=\footnotesize, anchor=north, legend columns=-1},
    y tick label style={/pgf/number format/.cd,%
          scaled y ticks = false,
          set thousands separator={},
          fixed
    },
        legend image code/.code={
        \draw [#1] (0cm,-0.1cm) rectangle (0.2cm,0.2cm);
    },
    ]
    
   \addplot+ table[x=Benchmark,y=256entries]{\robPerfBp};
    \addplot+ table[x=Benchmark,y=512entries]{\robPerfBp};
    \addplot+ table[x=Benchmark,y=1024entries]{\robPerfBp};
\end{axis}
\end{tikzpicture}
        \caption{\texttt{perfect-bp}: IPC rises with ROB on \texttt{bfs}/\texttt{cc}; \texttt{bisort}/\texttt{pagerank} gain modestly.}
    \end{subfigure}%
    ~ 
    \begin{subfigure}[t]{0.5\columnwidth}
        \centering
         \begin{tikzpicture}
    \begin{axis}[
    ybar,
    ylabel={IPC},
    y label style={at={(0.06,0.5)}},
    every x tick label/.append style={font=\footnotesize},
    width=\linewidth,
    height=3cm,
    x tick label style={rotate=0}, 
    symbolic x coords={
    bfs,bisort,cc,pagerank    },
    ymin=0,
    xtick=data,
    bar width=3,
    enlarge x limits=0.15,
    legend entries={},
        legend style={at={(0.4\columnwidth,1.2)}, anchor=north, legend columns=-1},
    y tick label style={/pgf/number format/.cd,%
          scaled y ticks = false,
          set thousands separator={},
          fixed
    },
    ]
    
   \addplot+ table[x=Benchmark,y=256entries]{\robPerfCache};
    \addplot+ table[x=Benchmark,y=512entries]{\robPerfCache};
    \addplot+ table[x=Benchmark,y=1024entries]{\robPerfCache};
\end{axis}
\end{tikzpicture}
        \caption{\texttt{perfect-cache}: IPC is nearly flat because memory is already ``perfect'' and squashes dominate.}
    \end{subfigure}
    \caption{\textbf{ROB scaling under isolated perfect modes (256/512/1024 entries).}
    \texttt{perfect-bp} benefits when fewer squashes expose load latency; \texttt{perfect-cache} is insensitive because wrong-path work dominates.
    Under joint \texttt{perfect}, IPC is likewise flat (not shown).}
    \label{fig:rob_effect}
\end{figure}

Under \texttt{perfect-bp}, \texttt{bfs} and \texttt{cc} gain IPC as a larger window absorbs exposed load latency.
Under \texttt{perfect-cache}, IPC is nearly flat because unresolved branches, not window depth, dominate; joint \texttt{perfect} is also insensitive.
A larger window can reduce measured coupling for some kernels by tolerating more exposed memory latency, but it does not eliminate SSL when the hot loop still couples an unpredictable branch with an irregular load.

\subsubsection{Upper-bound speedup}
\begin{figure*}[!ht]
\centering
\includegraphics[width=\textwidth,height=3.7cm,keepaspectratio=false]{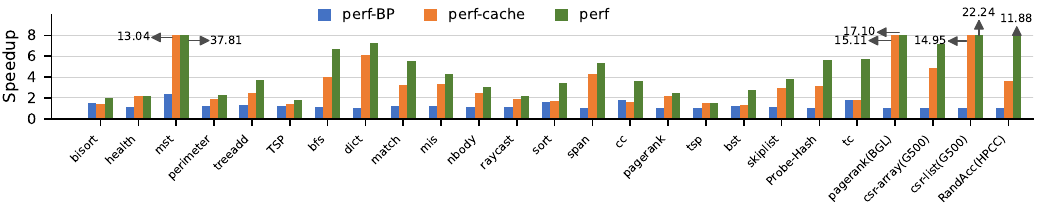}
\caption{\textbf{Speedup across simulation modes (screened kernels, normalized to \texttt{base}).}
Bars show \texttt{perfect-bp}, \texttt{perfect-cache}, and joint \texttt{perfect} speedup.
On coupled-stall kernels (\texttt{tc}, \texttt{mst}, \texttt{RandAcc}), the joint bar far exceeds either isolated mode; near-independent kernels (\texttt{health}, \texttt{tsp}) approach joint \texttt{perfect} from isolated modes.
Arrow labels give values above the axis limit.}
\label{fig:upper_bound_ipc}
\end{figure*}

The final hardware test asks how much the joint ceiling exceeds the isolated ceilings.
\Cref{fig:upper_bound_ipc} shows that \texttt{tc} gains only ${\sim}$1.8$\times$ from each isolated mode, yet reaches ${\sim}$5.8$\times$ under joint \texttt{perfect}.
\texttt{mst} is more extreme: ${\sim}$2.3$\times$ and ${\sim}$13$\times$ isolated gains versus ${\sim}$38$\times$ joint.
Once loads are fast, squashes occur more often per unit time; once branches are correct, outstanding misses accumulate behind the ROB head.
Thus, isolated modes reveal the remaining bottleneck, while the joint mode reveals headroom missed by independent-ceiling analysis.

%% file: discussion.tex
\subsection{SPEC CPU2017 benchmarks}
The kernel results in \Cref{sec:results} expose SSL in tight irregular loops.
SPEC CPU2017 is stricter: full applications can separate branch-bound and memory-bound phases, so aggregate MPKI and geomeans may hide short SSL-heavy intervals or overstate coupling when the event classes occur in different regions.

\begin{figure*}[!t]
\centering
\begin{subfigure}[t]{0.32\textwidth}
  \centering
  \includegraphics[width=\linewidth,height=3.55cm,keepaspectratio]{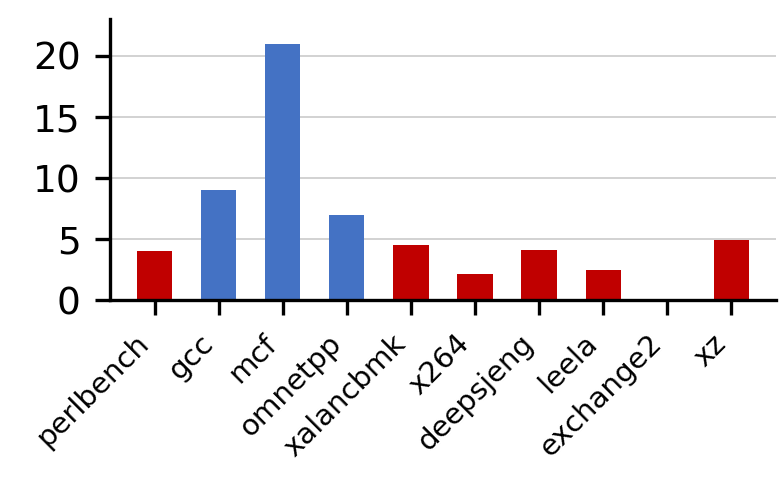}
  \caption{SSO per benchmark}
  \label{fig:spec_mpki}
\end{subfigure}\hfill
\begin{subfigure}[t]{0.44\textwidth}
  \centering
  \includegraphics[width=\linewidth,height=3.55cm,keepaspectratio]{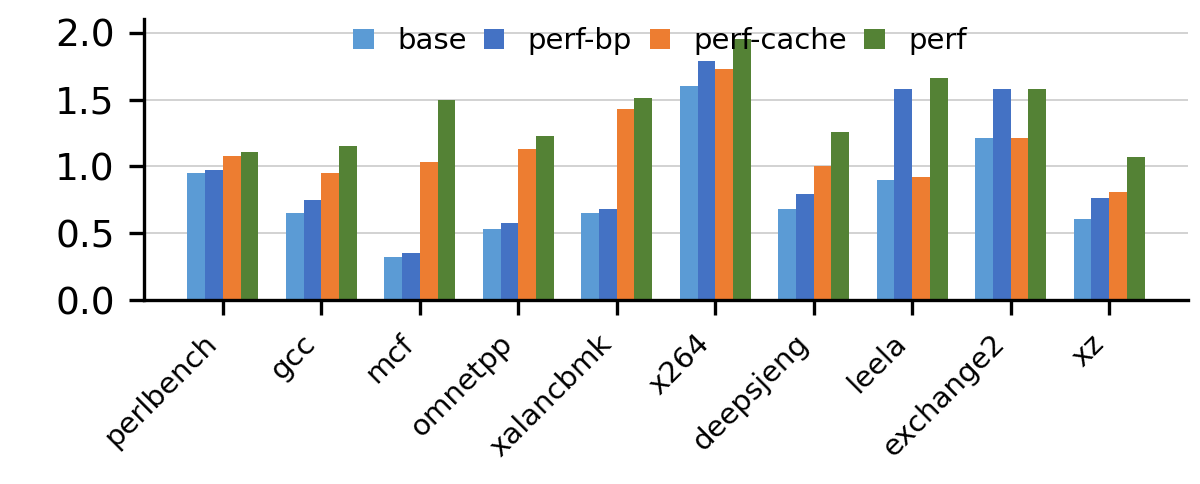}
  \caption{IPC under \texttt{base}, \texttt{perf-bp}, \texttt{perf-cache}, and \texttt{perf}}
  \label{fig:spec_ipc}
\end{subfigure}\hfill
\begin{subfigure}[t]{0.22\textwidth}
  \centering
  \includegraphics[width=\linewidth,height=3.55cm,keepaspectratio]{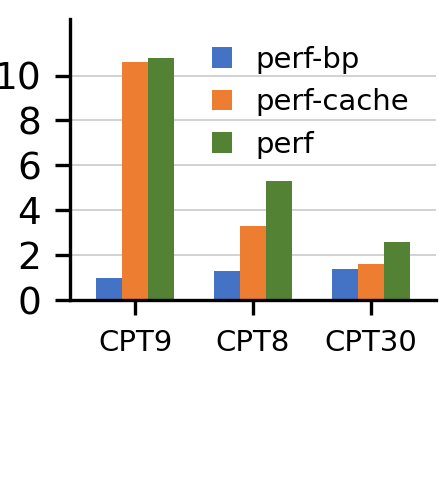}
  \caption{\texttt{mcf} checkpoint speedups}
  \label{fig:mcf_speedup}
\end{subfigure}
\caption{\textbf{SPEC CPU2017 int: (a) SSO screening, (b) IPC under different modes, and (c) \texttt{mcf} checkpoint speedups.}
Only \texttt{605.mcf\_s} is high-SSO/high-JSS in aggregate. Per-checkpoint speedups show why aggregation can hide SSL: CPT9 is cache-dominated (JSS $\approx$ 1.05), whereas CPT8 and CPT30 show coupled headroom (JSS $\approx$ 1.35 and 1.51).}
\label{fig:spec_overview}
\end{figure*}

We evaluate SPEC CPU2017 integer-speed checkpoints with the same gem5 configuration and perfect modes as the kernel study (\Cref{sec:methodology}).
Following \cite{weisse2019nda}, we aggregate 100 native checkpoints per benchmark, sampled with SMARTS \cite{smarts} and Lapidary-style weighting \cite{lapidary}.
\Cref{fig:spec_overview} reports SSO, IPC under the four modes, and representative \texttt{mcf} checkpoint speedups.
Most SPEC workloads have low SSO or near-independent aggregate behavior ($\mathrm{JSS}\approx 1$).

\texttt{605.mcf\_s} is the only SPEC benchmark in our runs that is high-SSO/high-JSS in aggregate ($\mathrm{JSS}\approx 1.4$).
CPT9 is cache-dominated (${\sim}$10.5$\times$ from \texttt{perfect-cache}, $\mathrm{JSS}\approx 1.05$), whereas CPT8 and CPT30 interleave pointer-chasing loads and data-dependent branches ($\mathrm{JSS}$ of 1.35 and 1.51).
SPEC-style SSL claims should therefore report phase- or checkpoint-level slices alongside aggregate geomeans.

\subsection{Recommended Evaluation Framework}
JSS is an evaluation diagnostic: mechanisms targeting branch prediction, caching, prefetching, or irregular pointer-chasing workloads should make branch--memory coupling explicit rather than relying on isolated ceilings.

\textbf{Conditional subsystem reporting.}
Evaluate branch predictors under both the baseline memory system and a stronger cache or \texttt{perfect-cache} configuration.
If prediction gain increases after cache misses are reduced, memory latency was masking branch value and isolated \texttt{perfect-bp} understates prediction payoff.
Conversely, evaluate prefetchers and caches under baseline and stronger or perfect branch prediction: if memory-side gain rises after prediction improves, branch squashes were masking memory value.

\textbf{Benchmark selection.}
Use SSO to screen and JSS to confirm coupling.
High-SSO/high-JSS kernels such as \texttt{tc}, \texttt{mst}, \texttt{probe-bst}, \texttt{bfs}, and \texttt{RandAcc} should appear in evaluations that claim to address coupled stalls.
Suite names alone are insufficient: pointer-heavy graph kernels often show high JSS, while array-based implementations can remain near independence.
For full applications, pair geomeans with phase-level slices, as \texttt{mcf} illustrates in \Cref{fig:spec_overview}.

\textbf{Design triage.}
SSO and JSS form a compact decision matrix (\Cref{tab:sso_jss_triage}).

\begin{table}[!t]
\centering
\caption{SSO/JSS design triage.}
\label{tab:sso_jss_triage}
\scriptsize
\begin{tblr}{
  width = \linewidth,
  colspec = {Q[l,m,0.18\linewidth] X[l,m] X[l,m]},
  row{1} = {font=\bfseries},
  hlines, vlines,
  stretch = 1.05,
  colsep = 3pt,
}
Metric profile & Diagnostic meaning & Recommended action \\
Low SSO & Branch and cache penalties are not simultaneously frequent. & Isolated branch/cache evaluation is usually sufficient. \\
\shortstack[l]{High SSO,\\Low JSS} & Both penalties occur, but are phase- or path-separated. & Use phase slicing or PC attribution before proposing a coupled design. \\
\shortstack[l]{High SSO,\\High JSS} & Penalties are frequent and mutually limiting in the same hot loops. & Evaluate coupled branch/memory mechanisms and report joint and conditional gains. \\
\end{tblr}
\end{table}

\textbf{Reporting checklist.}
Report isolated, joint, and conditional branch-after-cache and cache-after-branch gains together with JSS.
Modest isolated gain but high conditional gain points to a coupled path; $\mathrm{JSS}\approx 1$ supports conventional isolated ceilings.

\subsection{Scope and Interpretation}
These results are diagnostic upper bounds on coupled branch--memory opportunity, not speedup predictions for a deployable mechanism.
A high JSS value does not imply that a practical predictor or prefetcher can realize the full joint speedup; it identifies where isolated evaluation is likely to understate headroom.
High SSO should be confirmed with JSS because the two event types may occur in different phases or dynamic paths, especially in full applications where aggregate MPKI can mix separate branch-bound and memory-bound regions.

Absolute JSS may vary with core width, window size, cache hierarchy, predictor, prefetcher, and threading model.
Our ROB sensitivity study (256/512/1024 entries) shows that a larger window can reduce coupling for some workloads by tolerating more exposed memory latency, but it does not eliminate SSL when irregular loads and unpredictable branches remain mutually limiting in the same hot loop.
The broader recommendation is to measure joint and conditional gains rather than assume composability.

This paper is intentionally methodological.
Future work includes using SSO/JSS to select phases for coordinated branch--memory optimization, dynamically enabling coupled mechanisms in high-pressure regions, evaluating existing load-dependent predictors and control-flow-decoupling techniques with branch-after-cache and cache-after-branch measurements, and extending the study to multicore, shared-cache, and full-system settings.

%% file: Conclusion.tex
Perfect-branch and perfect-cache speedups should not be assumed to compose. Across 53 simulated kernels, 37 (70\%) have $\mathrm{JSS}>1$; 21 (40\%) exceed the independence product by more than 6\%; and workloads with $\mathrm{SSO}>20$ show $\mathrm{JSS}$ from 1.23 to 3.29. SSO is a low-cost screening metric, while JSS determines whether isolated upper bounds understate joint headroom. When JSS is near one, isolated evaluation is adequate; when it is elevated, branch and memory mechanisms should be evaluated jointly.

High-JSS cases concentrate in four recurring families, neighboring node access (NNA), hash table lookup (HTL), linked-structure traversal (LDST), and data-dependent modification (DDM), whose hot loops repeatedly pair irregular loads with data-dependent branches. Isolated perfect modes shift pressure rather than eliminate it: \texttt{perfect-bp} raises ROB occupancy by preserving memory-blocked work, while \texttt{perfect-cache} can raise squash events per cycle. Only joint \texttt{perfect} removes both limiters. SPEC CPU2017 reinforces the same lesson at application scale: most integer-speed workloads are near single-bottleneck after aggregation, but \texttt{605.mcf\_s} shows checkpoint-level coupling that aggregate geomeans can hide.

Practically, evaluations should report isolated, joint, and conditional branch-after-cache and cache-after-branch gains. Branch predictors should be measured under baseline and stronger or perfect-cache assumptions; prefetchers and cache mechanisms should be measured under baseline and stronger or perfect branch prediction. Benchmark suites should include high-SSO/high-JSS kernels when mechanisms claim to address coupled control and memory stalls.